\documentclass[traditabstract]{aa}
\usepackage{graphicx}

\begin{document}
\def\hr{\hbox{$^{\rm h}$}}                 
\def\fd{\hbox{$.\!\!^{\rm d}$}}            
\def\fhr{\hbox{$.\!\!^{\rm h}$}}           
\def\fdeg{\hbox{$.\!\!^\circ$}}            
\def\sec{\hbox{$^{\rm s}$}}                
\def\fsec{\hbox{$.\!\!^{\rm s}$}}          
\def\arcm{\hbox{$^\prime$}}                
\def\farcm{\hbox{$.\mkern-4mu^\prime$}}    
\def\farcs{\hbox{$.\!\!^{\prime\prime}$}}  
\def\day{\hbox{$^{\rm d}$}}                
\def\fday{\hbox{$.\!\!^{\rm d}$}}          
\def\per{\hbox{$^{\rm p}$}}                
\def\fper{\hbox{$.\!\!^{\rm P}$}}          
\def\mm{\hbox{$^{\rm m}$}}                 
\def\fmm{\hbox{$.\!\!^{\rm m}$}}           
\def\tim{\hbox{$\rm x$}}                   
\def\cd{$d^{\rm -1}$}
\def\Msol{$\rm{M_\odot}$}
\def\Rsol{$\rm{R_\odot}$}
\newcommand{\struut}{\rule[-2ex]{0ex}{5.2ex}}
\newcommand{\struutup}{\rule{0ex}{3.2ex}}
\newcommand{\struutdown}{\rule[-2ex]{0ex}{2ex}}
\newcommand{\mcol}[3]{\multicolumn{#1}{#2}{#3} }
\newcommand{\mco}{\mcol{8}{|c|}{ } \\}

\title{Multi-site, multi-year monitoring of the oscillating Algol-type eclipsing binary CT~Her\thanks 
{Based on photometric data collected at the observatories listed in Table~1 and spectra acquired at 
the NAO, Bulgaria, and Calar Alto Observatory, Spain. The Skinakas Observatory is a collaborative project 
of the University of Crete, the Foundation for Research and Technology -- Hellas, and the Max-Planck-Institut 
f\"ur Extraterrestrische Physik.},\thanks
{Tables~\ref{tb}, ~\ref{tv}, ~\ref{radvel} and ~\ref{octab} are only available in electronic form via anonymous 
ftp at cdsarc.u-strasbg.fr or at http://csdweb.u-strasbg.fr/cgi-bin/qcat?/A+A/XXX/YYY }
}

\author{P. Lampens           \inst{1}  
\and A. Strigachev           \inst{2}
\and S.-L. Kim               \inst{3}  
\and E. Rodr\'iguez          \inst{4}  
\and M.J. L\'opez-Gonz\'alez \inst{4}  
\and J. Vidal-Sa\'inz        \inst{5}  
\and D. Mkrtichian           \inst{6,7,8}
\and J.-R. Koo               \inst{3,9}
\and Y. B. Kang              \inst{3,9}
\and P. Van Cauteren         \inst{10,11}  
\and P. Wils                 \inst{11}  
\and Z. Kraicheva            \inst{2}
\and D. Dimitrov             \inst{2}
\and J. Southworth           \inst{12}
\and E. Garc\'ia Melendo     \inst{5}
\and J.M. G\'omez Forellad   \inst{5}  
}

\offprints{P. Lampens}

\institute{Koninklijke Sterrenwacht van Belgi\"e, Ringlaan 3, 1180 Brussel, Belgium \\
  \email{Patricia.Lampens@oma.be}
\and Institute of Astronomy and National Astronomical Observatory, Bulgarian Academy of Sciences, 72 Tsarigradsko Shosse Blvd., 1784 Sofia, Bulgaria 
\and Korea Astronomy and Space Science Institute (KASI), Daejeon 305-348, Korea
\and Instituto de Astrof\'{\i}sica de Andaluc\'{\i}a, CSIC, P.O. Box 3034, E-18080 Granada, Spain
\and Grup d'Estudis Astronomics, Apdo. 9481, 08080 Barcelona, Spain
\and Astrophysical Research Center for the Structure and Evolution of the Cosmos, Sejong University, Seoul, Korea
\and Astronomical Observatory, Odessa National University, Odessa, 650014 Ukraine
\and Crimean Astrophysical Observatory, Nauchny,  98409 Crimea,  Ukraine
\and Department of Astronomy and Space Science, Chungnam National University, Daejeon, 305-764, Korea
\and Beersel Hills Observatory (BHO), Beersel, Belgium
\and Vereniging Voor Sterrenkunde (VVS), Oostmeers 122 C, 8000 Brugge, Belgium
\and Astrophysics Group, Keele University, Newcastle-Under-Lyme ST5 5BG, United Kingdom 
}
\date{Received \dots; accepted \dots}

\titlerunning{The oscillating Algol-type eclipsing binary CT~Her}
\authorrunning{P.~Lampens et al.}

\abstract
{We present the results of a multi-site photometric campaign carried out in 2004-2008 for the Algol-type
eclipsing binary system CT~Her, the primary component of which shows $\delta$ Scuti-type oscillations. Our
data consist of differential light curves collected in the filters $B$ and $V$ which have been analysed using
the method of Wilson-Devinney ({\sc Phoebe}). After identification of an adequate binary model and removal 
of the best-matching light curve solution, we performed a Fourier analysis of the residual $B$ and $V$ light curves 
to investigate the pulsational behaviour. { We confirm the presence of rapid pulsations with a main
period of 27.2 min.} {Up to eight} significant frequencies with semi-amplitudes in the 
range 3 to 1 mmag were detected, all of which lie in the frequency range 43.5-53.5 \cd. This 
result is independent from the choice of the primary's effective temperature (8200 or 8700 K) { since 
the light curve models of the binary are very similar in both cases}. 
This is {yet another case of a complex frequency spectrum} observed for an accreting $\delta$ Scuti-type star {(after Y~Cam)}. 
In addition, we demonstrate that the amplitudes of {several} pulsation frequencies show evidence of variability 
on time scales as short as 1-2 years, perhaps even less. Moreover, our analysis takes into account some recently acquired 
spectra, from which we obtained the corresponding radial velocities for the years 2007--2009. Investigation of the O-C diagram 
shows that further monitoring of the epochs of eclipse minima of CT~Her {will cast} a new light on the evolution of its 
orbital period.} 


\maketitle

\section{Introduction}

Asteroseismology aims to understand the pulsation physics in order to probe the interior parts of all kinds 
of stars. Among the classical pulsators, $\delta$ Scuti stars are rather common and are located at the 
intersection of the Cepheid instability strip and the main sequence. They pulsate primarily in low overtone 
radial and non-radial acoustic modes with periods between 30 min and 6 hr. Some may also pulsate in gravity 
modes (\cite{kur}). Their excitation mechanism, i.e. the $\kappa$-mechanism active in the partial ionisation 
zones of He~I and He~II, is well-understood. However, still now, unknown amplitude limiting and mode selection 
mechanisms are operating in these pulsators. Only a fraction of the theoretically predicted modes are 
observed, which results in (too) many free parameters in the pulsation models. It is {indeed} primordial 
to collect knowledge on the fundamental stellar properties of the pulsator independently. 

One way is to study the pulsating components of binary (multiple) systems in great detail, since binary 
(multiple) systems with well-characterized components supply additional constraints for a more reliable 
modelling. The positions in the H-R diagram and therefore the components' evolutionary statuses can be
more accurately determined than in the case of single stars, e.g. for those $\delta$ Scuti stars
that are at the (very) end of their H-core burning phase. The complexity of having to deal with 
additional components does generally not weigh against the scientific return, although observations 
spread over even longer time-scales may be necessary to disentangle both phenomena.

About seventy percent of all stars of the Solar neighbourhood are members of binary or multiple systems 
(67\% for G-M stars, \cite{may}; ~75\% for O-B stars, \cite{mas}). Yet, these facts are usually ignored 
in the study of stellar pulsation. There is, however, strong evidence that duplicity affects the pulsation 
properties in specific cases (e.g. the eccentric binaries HD~177863 (\cite{dc}) and HD~209295 (\cite{ha})). 
This is also predicted from a theoretical point-of-view (\cite{wit99}). It is thus essential to understand 
the possible link(s) between binarity and pulsation(s) and to observe whether or not - under given circumstances - 
the internal structure and the pulsating properties of such stars might be different from those of the 
single pulsators (\cite{lam06}).

Detached and semi-detached eclipsing binaries (EBs) are particularly powerful tools in astrophysics: their 
accurate observation enables to derive the fundamental properties (masses, radii, luminosities) of each 
component. EBs also provide the component's effective temperatures, as well as the distance if stellar 
atmosphere models are used (\cite{mac}).

We present a detailed photometric study of the Algol-type binary CT~Her, an eclipsing binary of mag 11-12 and 
spectral type A3V+[G3IV] with an orbital period of 1.7863748 days (\cite{sam}). {Radial velocity measurements have also been collected, but due to its faintness, high-quality spectra are not easily acquired.} CT~Her belongs to the { group} of oscillating Algol-type (oEA) stars comprising $\approx$ 35 known members. {Up to now,} a few members of this { group} only have been investigated thoroughly: RZ~Cas has been observed photometrically as well as spectroscopically over almost a decade (\cite{leh04}, Rodr\'{\i}guez et al.\ \cite{rod04,soy06,leh08}) {while a very detailed photometric study of Y~Cam was recently finished (\cite{rod10}). TW~Dra is another case for which a large spectroscopic observational effort was carried out (\cite{leh09,tka10}).} 
{Its (O-C) diagram shows a slow orbital evolution (\cite{kre}).} The primary component displays oscillations of type $\delta$ Scuti with a (main) pulsation period of 0.46 hr ($\approx$ 28~min) and a total amplitude of about 0.02 mag (Kim et al.\ \cite{kim}). 
{Among the currently known oEA stars, it has one of the shortest orbital periods and the highest ratio $P_{\rm orb}/P_{\rm puls}$ (about 95).}

The oEA stars are former secondaries of evolved, semi-detached eclipsing binaries which are (still) undergoing 
mass transfer and form a recently detected { subclass} of pulsators close to the main sequence (Mkrtichian et al.\, 
\cite{mkr1,mkr2}). 
Searches for new oscillating Algols have been performed by Kim et al.\, (\cite{kim1,kim2}) and Mkrtichian et al.\, 
(\cite{mkr3,mkr4}). \cite{pi07} and \cite{mi08} looked for them in the ASAS-3 and the OGLE-II public databases. 
A search using the NSVS database has also been on-going at the Institute of Astronomy of the Bulgarian Academy of 
Sciences (Dimitrov et al.\, \cite{dd08a,dd08b,dd09a,dd09b}). Their general characteristics 
and pulsational properties have been summarized by Mkrtichian et al.\ (\cite{mkr3}). 
Such oEA stars 
are, indeed, {excellent laboratories for investigating the effects of mass accretion events as well as of tides unto 
the pulsation properties. Changes {of these properties (amplitudes, modes and/or phases)} due to mass accretion episodes
have been observed} (e.g. in RZ~Cas where strong modal amplitude variations followed an abrupt change of its orbital period; Rodr\'{\i}guez et al.\ \cite{rod04}, Mkrtichian et al.\ \cite{mkr3}). 
Their pulsational frequencies could be 
tidally split (as in KW~Aur, \cite{fit76}) and/or there may be some coupling between the pulsation and the orbital 
frequency due to some resonance mechanism (Mkrtichian et al.\ \cite{mkr3}). Their very different evolutionary history 
is  a challenge for stellar evolution modelling. For example, in order to reproduce the gainers 
of RZ~Cas, KO~Aql and S~Equ through conservative binary evolution, large initial mass ratios (typically $>$ 3) would be 
necessary. But the corresponding high mass loss rates in the beginning of the Roche Lobe overflow stage result in radii 
larger than the Roche radii. Hence, a non-conservative approach is needed, in which processes are considered that enhance 
the period without losing too much mass (\cite{dgr}). Such studies have not yet been attempted.  
Detailed and multi-year studies of more oEA stars are essential in order to provide solid grounds for their (future) asteroseismic modelling.\\

\section{Observations and data reduction}

\subsection{The multi-site campaign}

High-precision light curves of CT~Her (GSC~01509-1142, $V=11.347,B-V=0.203$) were collected in the framework of a 5-yr 
long multi-site campaign set up with the purpose to study its pulsational behaviour. A logbook of the observations is 
given in Table~\ref{tnight}. We observed in a differential mode from late spring till late summer of the years 
2004-2008. The original observations which led to its discovery as an oEA star (Kim et al.\ \cite{kim}) were also included. The following comparison stars were used: C1=GSC~01509-1140 ($V=10.951,B-V=1.414$); C2=GSC~01509-1052 ($V=11.405,B-V=0.779$); C2'=GSC~01509-1130 
($V=12.02$); C3=GSC~01509-0901 ($V=12.20,B-V=0.78$) and C4=GSC~01509-1090 ($V=10.75,B-V=0.48$). 

The characteristics of the various CCD cameras are the following ones:\\ 
-- at the Skinakas Observatory of the University of Crete, Greece. 
The camera is a Photometrics $1024\times1024$ with a SITe~SI003B chip of grade~1 and a pixel size of $24~\mu$m corresponding to a scale of 
$0.5\arcsec$ on the sky. The field-of-view is $8.5\arcmin\times8.5\arcmin$ (\cite{pa}).\\
-- at the Sobaeksan Optical Astronomy Observatory of the Korea Astronomy and Space Science Institute (KASI), South-Korea. The camera has a $2048\times2048$ SITe chip with a pixel size of $24~\mu$m corresponding to a scale of $0.60\arcsec$ on the sky. The field-of-view is 
$20.5\arcmin\times20.5\arcmin$. \\
-- at the Observatory of Mt. Lemmon, Arizona, operated by KASI. The camera has a $2084\times2084$ Kodak chip with a pixel size of $24~\mu$m corresponding to a scale of $0.64\arcsec$ on the sky. The field-of-view is $22.2\arcmin\times22.2\arcmin$ large.\\
-- at the Observatorio de Monegrillo\footnote{http://www.astrogea.org/jvidal/index.html}, North of Spain. The camera is a SX Starlight CCD with a Sony ICX027BL chip (cooled to about -25\degr C) and a pixel size of $12.7~\mu\times16.6~\mu$ corresponding to $1.80\arcsec\times1.38\arcsec$. The field-of-view covered a sky region of $11.5\arcmin\times7.7\arcmin$. 
The reduction was done using a software package called LAIA (Laboratory for Astronomical Image Analysis) developed by Joan A. Cano\footnote{http://www.astrogea.org/soft/laia/laia.htm}. \\
-- at the Beersel Hils Observatory, Belgium. The camera is a SBIG ST10XMe with a chip of grade~1 and a pixel size of $6.8~\mu$m corresponding to a scale of $1.43\arcsec$ on the sky (in 2$\times$2 binning). The field-of-view is $17.5\arcmin\times26\arcmin$ on the sky. 
The reduction was done using the Mira AP (v.7) package\footnote{The Mira software is a registered trademark of Mirametrics, Inc., http://www.mirametrics.com/index.htm}.\\
-- at the Observatorio Sierra Nevada, South of Spain. The camera has a 2k$\times$2k chip with a pixel size of $13.5~\mu$m 
corresponding to a scale of $0.23\arcsec$ on the sky (2$\times$2 binning). The field-of-view is $7.92\arcmin\times7.92\arcmin$ large.\\
-- at the National Astronomical Observatory (NAO) Rozhen, Bulgaria.
The CCD is a VersArray 1330B with a $1340\times1300$ E2V CCD36-40 chip of grade~2 and a pixel size of $20~\mu$m corresponding to a scale of $0.258\arcsec$ on the sky. The field-of-view is $5.76\arcmin\times5.59\arcmin$ large.

\begin{table*}[t]
\caption[]{\label{tnight} {Campaigns and instrumentation. Total number of useful observations and hours.}}
\begin{tabular}{llllllllll}
\hline
Year & Site          & Country  & Tel.   & Period (\& Obs.)   & Nights & No. & Hours & Filter & CODE  \\
\hline                                                            	                                      
2004 & Skinakas      & Crete    & 1.3-m  & Jun-July (AS)      & 4 & 573   & 17 & $B$  &  SKB1   \\ 
2004 & Sobaeksan     & Korea    & 0.61-m & March (SLK)        & 3 & 379   & 10 & $B$  &  KRB1   \\ 
2005 & Skinakas      & Crete    & 1.3-m  & May-Aug (AS)       & 6 & 1091  & 29 & $B$  &  SKB2   \\ 
2005 & Mt. Lemmon    & Arizona  & 1.0-m  & Jun-July (SLK)     & 5 & 643   & 21 & $B$  &  KRB2   \\ 
2005 & Monegrillo    & Spain    & 0.4-m  & Jun-July (JV)      & 15& 1208  & 67 & $V$  &  SPV1   \\ 
2005 & Beersel       & Belgium  & 0.4-m  & Jun-July (PVC)     & 5 & 430   &  6 & $B$  &  BHOB   \\ 
2005 & Beersel       & Belgium  & 0.4-m  & Jun-July (PVC)     & 2 & 517   &  6 & $V$  &  BHOV   \\ 
2006 & Skinakas      & Crete    & 1.3-m  & Jun-July (AS)      & 9 & 1427  & 35 & $B$  &  SKB3   \\ 
2006 & Sierra Nevada & Spain    & 1.5-m  & Aug (ER+MLG)       & 9 & 835   & 18 & $B$  &  SPB1   \\ 
2007 & Sierra Nevada & Spain    & 1.5-m  & Mar-May(ER+MLG)    & 8 & 2158  & 18 & $B$  &  SPB2   \\ 
2007 & Beersel       & Belgium  & 0.4-m  & March (PVC)        & 1 & 83    &  3 & $V$  &  BHOV   \\ 
2008 & NAO Rozhen    & Bulgaria & 2.0-m  & June-July (AS)     & 2 & 358   & 12 & $B$  &  ROZB   \\
2008 & Monegrillo    & Spain    & 0.4-m  & June-July (JV)     & 3 & 372   & 10 & $V$  &  SPV2   \\
\hline
2004-08 &  All       &  ---	& ---	  &  ---     & ---    & 2180	  & 86 & $V$  &  ALL\\
2004-08 &  All       &  ---	& ---	  &  ---     & ---    & 7960	  & 166 & $B$  &  ALL\\
\hline
\end{tabular}
\end{table*}

Photoelectric photometric data on CT~Her were also acquired in the four Str\"omgren passbands uvby at the Observatorio Sierra Nevada between 2008 March, 31 and 2008 April, 13. 
The comparison stars were {K1}~=~HD 145122 ($V=6.13$), {K2}~=~HD 146101 ($V=8.24$) and {K3}~=~HD 145549 ($V=8.16$)). The observations were carried out in the sequence Sky, {K1}, {K2}, CT~Her, AO~Ser, {K1}, {K3}, CT~Her, AO~Ser\dots because {both} oEA stars are close enough to one another on the sky. Unluckily, the orbital period was not fully covered { during the additional observations}. 

\subsection{Observational technique}

CT~Her was observed using the standard (Johnson) filters $B$ (mostly) and $V$ (less frequently). At the Beersel Hills Observatory 
(BHO), we follow the specifications for the filters of Bessell (1995). C1 is the principal comparison 
star which was commonly used at every observatory. We adopted the filter $B$ as the main filter because of its higher 
signal-to-noise ratio and the larger amplitudes of pulsation expected. Typical exposure times for the $B$ filter were set 
between 15 s (e.g. Sierra Nevada) and 60 s (e.g. Skinakas). Typical exposures for the $V$ filter were 20-30 s (e.g. BHO and 
Skinakas). All observers followed the standard calibration procedure: a set of biases (resp. darks) was taken regularly during  each night and a set of 5 to 6 flat-fields per filter was obtained during evening and/or morning twilights. 

As an example, we describe the full reduction procedure used for images collected at the Skinakas Observatory. 
All the primary reduction steps were performed using standard ESO-MIDAS routines. The frames were processed as follows: subtraction of the residual bias pattern using a median master bias frame, flat-fielding using a median master flat-field frame, and median cosmic ray cleaning. Since the field is not crowded, the technique of aperture photometry was applied to extract the differential magnitudes. The fixed aperture photometry was performed using DAOPHOT (\cite{st}). CT~Her and the comparison stars C1, C2' and C3 were measured using an aperture size as close as possible to the value providing the highest signal-to-noise ratio (\cite{str}). The data consist of differential photometry of the variable 
star in the sense (CT~Her - C1) and of the check stars in the sense (C2'- C1) and (C3 - C1).


CT~Her was also observed in the $BVR$ filters together with standard fields (\cite{lan}). We performed all-sky 
photometry up to an airmass of 2 at the Skinakas Observatory on Aug. 2005, 1. The coordinates, calibrated magnitudes 
and colours of the { target and comparison} stars are { listed} in Table~\ref{tcalib}.

\begin{table*}[t]
\caption[]{\label{tcalib} Coordinates and standard system calibrated magnitudes with their errors for the target and comparison stars} 
\begin{tabular}{llccccccc}
\hline
Ident & GSC & RA           & DEC         & $B$           & $V$          & $R$          & $B-V$       & $V-R$       \\ 
\hline                                                                                          
CT Her & GSC01509-1142 & 16 20 26.57 & +18 27 16.9 & 11.33$\pm$0.03 & 11.12$\pm$0.03 & 11.00$\pm$0.03 & 0.21$\pm$0.04 & 0.12$\pm$0.04 \\ 
C1$^a$ & GSC01509-1140 & 16 20 33.89 & +18 27 18.5 & 12.06$\pm$0.03 & 10.80$\pm$0.03 & 10.11$\pm$0.03 & 1.26$\pm$0.04 & 0.69$\pm$0.04 \\ 
C2'$^b$& GSC01509-1130 & 16 20 18.61 & +18 27 32.6 & 13.41$\pm$0.03 & 12.40$\pm$0.03 & 11.88$\pm$0.03 & 1.01$\pm$0.04 & 0.52$\pm$0.04 \\ 
C3     & GSC01509-0901 & 16 20 43.00 & +18 30 53.2 & 13.18$\pm$0.03 & 12.67$\pm$0.03 & 12.38$\pm$0.03 & 0.51$\pm$0.04 & 0.29$\pm$0.04 \\ 
\hline
\end{tabular}

$^{a}$ \small {C1 was also used by Kim et al.\ (\cite{kim})} \\
$^{b}$ \small {C2' is not the same star as C2 (Kim et al.\ \cite{kim})} \\

\end{table*}

\section{Photometric data -- time series}\label{Pdata}

The influence of interstellar and atmospheric extinction may be crucial when constructing light curves of eclipsing binaries (\cite{prsa1}). This is generally the case when matching data from different sites and campaigns but it may also be valid when using data from a single site. Observations from different nights and/or sites do usually not match - their mean level may be shifted, and there may also a correlation with airmass during some nights. In this case, {differential corrections associated with the second-order coefficient of the atmospheric extinction (k'')} were needed since the colours of the comparison star (C1) and the variable star are not the same (see Table~2). These corrections were mainly needed for the data {acquired} in the filter $B$ because of the dependence in wavelength and the greater inhomogeneity of the $B$-data sets collected over {a longer} period and at a larger number of { sites} equipped differently. 


{To correct for this influence, we first computed the magnitude of CT~Her based on the magnitude of C1 
in ~Table~\ref{tcalib}} and the differential values (CT~Her - C1). Next, we adopted a preliminary model light curve based on known and fitted parameters for the $V$ light curve obtained during the 2005 \& 2008 runs at the Observatory of Monegrillo, Spain (cf.~Table~\ref{tnight}). This was possible because the $V$ light curve is less dependent on these corrections. {In general, a { clear dependence of the residuals} in magnitude difference with the airmass per night was found, which we modelled using linear regression. The correction involved the zero-point shift due to interstellar and atmospheric extinction and the slope due to the term associated to the second-order atmospheric extinction. At any given site with the same instrument and filter, this slope was assumed constant. Where relevant, we applied the correction. Remark that the effect is systematically smaller in the $V$-band than in the $B$-band.} The resulting differential magnitudes of every night were concatenated per filter to produce the overall corrected $B$ and $V$ data sets. These data are presented in Tables~\ref{tb} and ~\ref{tv}. 
The { full} tables are available in electronic form only. Table~\ref{tnight} also lists the numbers of observations collected at each observatory. In the subsequent analysis, more than 7900 and 2100 data points, respectively in the filters $B$ and $V$, were {treated}. 

\begin{table}[t]
\caption[]{\label{tb} CT~Her $B$ light curve (first three lines, table available in electronic form) }
\begin{tabular}{lccccc}
\hline
HJD           &  $B$         & Code & Year   \\ 
\hline                                                                                                
2453954.47391 & 11.460156 & SPB1 & 2006 \\	      
2453954.47529 & 11.462120 & SPB1 & 2006 \\	      
2453954.47609 & 11.460165 & SPB1 & 2006 \\	      
\dots         &           &      &      \\
\hline
\end{tabular}
\end{table}

\begin{table}[t]
\caption[]{\label{tv} CT~Her $V$ light curve (first three lines, table available in electronic form) }
\begin{tabular}{lccccc}
\hline
HJD           & $V$       & Code & Year  \\ 
\hline                                                                                                
2453557.38195 & 11.953100 & SPV1 & 2005 \\
2453557.38406 & 11.938400 & SPV1 & 2005 \\
2453557.38617 & 11.921600 & SPV1 & 2005 \\
\dots         &           &      &      \\
\hline
\end{tabular}
\end{table}


\section{Simultaneous modelling of the light curves}\label{LCmodel}

The $B$- and $V$-data sets { were used} to plot the respective light curves phased against the orbital period of
1.7863748 days (\cite{sam}). Using this ephemeris, we redetermined the (single) epoch of primary eclipse observed 
in the filter $V$ and obtained the improved value of $HJD_0 = 2442522.93270 \pm 0.00006$ days. 
The phased light curve in the filter $V$ is almost complete (Fig.~1). However, we lack most of the primary minimum in the filter 
$B$ { (Fig.~2)}. Our current objective is to compute an adequate model from both curves in order to remove the geometric and 
photometric effects caused by the orbital motion and to extract the smallest possible residuals for a subsequent frequency-analysis. 
Concise reports of previous analyses (Lampens et al.\ \cite{lam08a,lam08b}) did not include all of the above mentioned data sets. 
In particular, the full coverage of the primary eclipse was not (yet) achieved in the filter $V$. The data collected in 2008-2009 
were thus mostly aimed at filling the gaps in orbital coverage.

\begin{figure}[t]
\label{Fig1a}
\resizebox{8cm}{!}{\includegraphics*{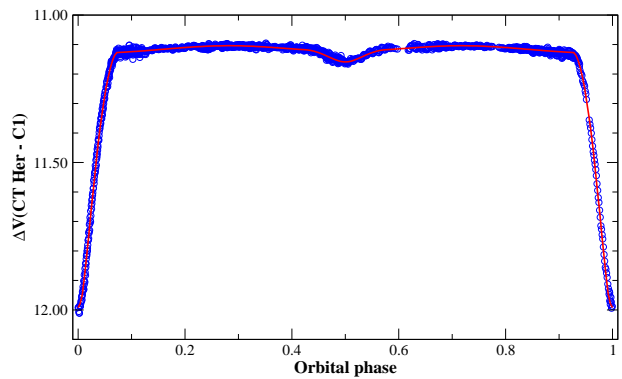}}
\caption[]{Observed and theoretical light curve in the filter $V$. Observations are shown as circles. The {solid} line represents the model.}
\end{figure}

\begin{figure}[]
\label{Fig1b}
\resizebox{8cm}{!}{\includegraphics*{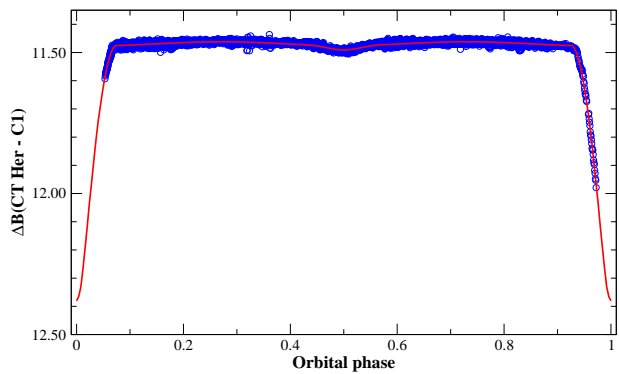}}           
\caption[]{Observed and theoretical light curve in the filter $B$. Same legend as in Fig.~1.}
\end{figure}

\begin{figure}[]
\label{Fig2}
\resizebox{8cm}{!}{\includegraphics*{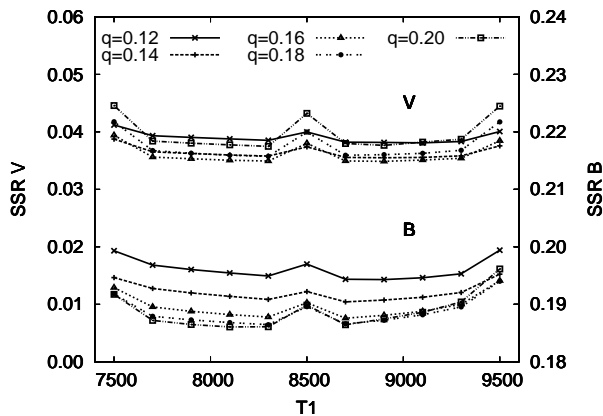}}           
\caption[]{$\chi{^2}$ as a function of $T_1$ and $q$}
\end{figure}

We performed a simultaneous modelling of the $B$ and $V$ light curves of CT~Her using the light-curve fitting programme 
{\sc Phoebe}, version 0.31a, in Mode 5, which corresponds to a semi-detached binary configuration in which the secondary 
component fills its limiting lobe, as required for the oEA stars. {\sc Phoebe} (\cite{prsa2}) is a package which
enables to compute models of eclipsing binaries based on observed photometric and radial velocity data. It relies on 
the 2003-version of the widely used Wilson-Devinney code (\cite{wd1}, Wilson~\cite{wd2,wd3}).
First, we employed the almost complete and homogeneous $V$-light curve obtained by one of us (JVS) during the years 2005 and 
2008 to compute an { initial} model; next, using this initial model, we corrected the individual {$B$ and $V$ data sets (where 
needed) for the differential effects of atmospheric extinction} as explained in Sect.\ref{Pdata}. 
Only then did we compute a best-fitting solution for the combined data series. 

Since CT~Her {is classified as an A3V star} (\cite{sk}), we initially adopted 8700~K as the surface temperature of the primary component. 
Except for the orbital ephemeris, {all other} parameters were set as adjustable parameters during the minimization procedure: 
this concerns the secondary star's surface temperature, $T_2$, the inclination, $i$, the mass ratio, $q$, the dimensionless 
potential, $\Omega_1$, and the fractional luminosities of the primary component, $L_1$, in $B$- and $V$-light (6 free parameters). 
We adopted the { roughly} determined absolute photometric and geometric elements (\cite{sv}) as starting values (e.g. 
$i=82\degr$ for the inclination and $q=0.27$ for the mass ratio), together with an estimate of the secondary component's surface 
temperature, $T_2 = 5800$ K, corresponding to a spectral type of G3IV. Other parameters such as the gravity darkening coefficients $g_1$ 
and $g_2$ and the albedos $A_1$ and $A_2$ were set to the theoretical values corresponding to a radiative atmosphere ($g_1=1.0$ and 
$A_1=1.0$) in the case of the primary component and to a convective atmosphere ($g_2=0.32$ and $A_2=0.5$, \cite{ru}) in the case of 
the secondary component. The limb darkening coefficients in the $B$- and $V$-bands were taken from the tables by \cite{vh}. In all 
the runs, convergence was achieved after only two or three iterations using the differential correction method.

By scanning the distribution of the function to be minimized, $\chi{^2}$, as a function of both $T_1$ and the mass ratio, $q$, we 
found two regions of possible solutions: one region
{near} $T_1=8700$ K and another one {near} $T_1=8200$ K. Fig.~3 illustrates the evolution of $\chi{^2}$ 
as a function of the parameters $T_1$ and $q$. Thus, we considered two { equivalent solutions} in the subsequent analysis: 
{ one solution associated with $T_1=8700$~K (model~71) and 
one associated with $T_1=8200$~K (model~72), with their corresponding values of $T_2$ and $q$}. 
The latter value of $T_1$ corresponds to a spectral type of A5V and to a colour index ($B-V$)$=0.15$ assuming zero reddening (cf. also the observed 
colour index ($B-V$) in Table~\ref{tcalib}).
The resulting parameter values and their formal uncertainties are listed in Table~\ref{t71}. 
Both models fit the observed $B$ and $V$ light curves very well, as evidenced by the small values of the $\chi{^2}$ function and the 1-2\%-level 
scatters of their residual data sets. The synthetic (and observed) light curves 
corresponding to the first model are illustrated by Figs.~1 and~2. The synthetic light curves corresponding to the alternative model 
are {indistinguishable}. 

The main difference between both models is the surface effective temperature of the secondary component, $T_2$, which is shifted 
by {about} 200 K. The largest changes with respect to the initial values are found in the parameters $T_2$ and $q$: both are smaller 
than their adopted first guesses (respectively 5800~K and 0.27). The mass ratio is significantly smaller than previously assumed 
(i.e. $q< 0.20$). 
Nonetheless, even though the components' effective temperatures are different, both models { are obviously similar} since their 
{ derived} physical parameters { lie} very close to one another (cf. the masses, gravities and radii of Table~\ref{phys}). 

From here on, even though the values of the $\chi{^2}$ function are slightly smaller in the former case (cf. Table~\ref{t71}),
we will adopt the solution derived with $T_1=8200$~K (model~72), the reason being that we obtained a better fit with this temperature 
to a few observed regions of the spectrum of CT~Her (Sect.~\ref{spectro}). We will { furthermore} show that 
the frequency-analysis of the corresponding residuals is {\it independent} of the choice of either one of the two proposed models 
(Sect.~\ref{freq}). 

\begin{figure}[t]
\label{Fig3a}
\centering
\resizebox{7.5cm}{!}{\includegraphics*{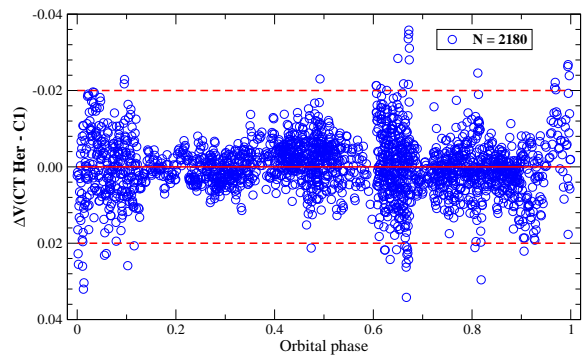}}           
\caption[]{Residual light curve phased against P$_{orb}$ (filter $V$).}
\end{figure}

\begin{figure}[]
\label{Fig3b}
\centering
\resizebox{7.5cm}{!}{\includegraphics*{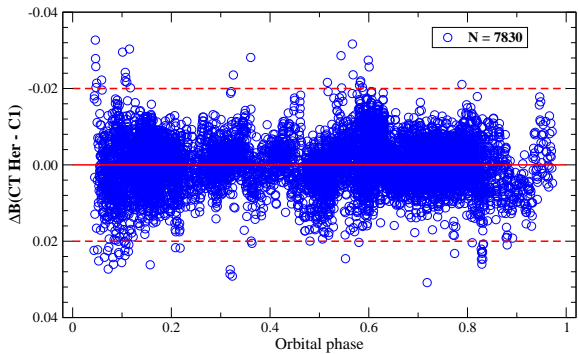}}           
\caption[]{Residual light curve phased against P$_{orb}$ (filter $B$).}
\end{figure}

\begin{table}
\caption{Parameters of the simultaneous {light curve solutions for CT~Her, including their formal error}.} 
\label{t71}
\begin{tabular}{l@{}c@{}c@{}c@{}c@{}}
\hline
\mcol{1}{l}{Parameter} & \mcol{2}{c}{Model~71} & \mcol{2}{c}{Model~72} \struutup\\
\mcol{1}{l}{} & \mcol{1}{c}{ Filter $B$ } & \mcol{1}{c}{Filter $V$ } & \mcol{1}{c}{Filter $B$ } & \mcol{1}{c}{Filter $V$} \struutdown\\
\hline
$i$ $(\degr)$       & \mcol{2}{|c}{81.95     $\pm$ 0.01     } & \mcol{2}{|c}{81.75     $\pm$ 0.01   } \struutup \\
$q$                 & \mcol{2}{|c}{0.1413    $\pm$ 0.0004   } & \mcol{2}{|c}{0.1456    $\pm$ 0.0003 } \\
$T_{1}$ (K)         & \mcol{2}{|c}{8700$^{a}$               } & \mcol{2}{|c}{8200$^{a}$             }  \\   
$T_{2}$ (K)         & \mcol{2}{|c}{4651      $\pm$ 7        } & \mcol{2}{|c}{4489      $\pm$ 7      }  \\
$\Omega_{1}$        & \mcol{2}{|c}{4.406     $\pm$ 0.006    } & \mcol{2}{|c}{4.426     $\pm$ 0.006  }  \\
$\Omega_{2}$ $^{c}$ & \mcol{2}{|c}{2.079                    } & \mcol{2}{|c}{2.091                  }  \\ 
$[{L_{1} \slash (L_{1} + L_{2})}]$ & 0.969 & 0.932           &  \mcol{1}{|c}{0.970}   & 0.935\\
$[{L_{2} \slash (L_{1} + L_{2})}]$ & 0.031 & 0.068           &  \mcol{1}{|c}{0.030}   & 0.065 \\
$g_1$               & \mcol{2}{|c}{1.0$^a$                  } & \mcol{2}{|c}{1.0$^a$      } \\
$g_2$               & \mcol{2}{|c}{0.32$^a$                 } & \mcol{2}{|c}{0.32$^a$     } \\
$A_1$               & \mcol{2}{|c}{1.0$^a$                  } & \mcol{2}{|c}{1.0$^a$      } \\
$A_2$               & \mcol{2}{|c}{0.5$^a$                  } & \mcol{2}{|c}{0.5$^a$      } \\
$x_{1}$$^{c}$       & \mcol{1}{|c}{0.561$^b$}         & 0.476$^b$          &  \mcol{1}{|c}{0.585$^b$}    & 0.505$^b$\\
$x_{2}$$^{c}$       & \mcol{1}{|c}{0.934$^b$}         & 0.785$^b$          &  \mcol{1}{|c}{0.962$^b$}    & 0.811$^b$ \struutdown\\
\hline
No. data     & \mcol{1}{|c}{7960}   & 2180  & \mcol{1}{|c}{7960}   & 2180 \struutup\\
$\chi^2$    & \mcol{1}{|c}{0.0160} & 0.0055 & \mcol{1}{|c}{0.0163} & 0.0056 \struutdown\\
\hline
\end{tabular}

$^{a}$ \small {adopted value} \\
$^{b}$ \small {Van Hamme (1993) tables} \\
$^{c}$  
$x_{1}$, $x_{2}$ are the limb darkening coefficients. $\Omega_2$ is the dimensionless potential of the secondary component.\\ 

\end{table} 

%

\section{Frequency analysis of the residual light curves}\label{freq} 

The light curve model with $T_1 = 8200$ K was subsequently subtracted from the original light curves of CT~Her 
to search for short-period pulsations in the residual data. The {phased} residual light curves are plotted 
in Figs.~4 and~5, 
respectively in the filters $V$ and $B$. The adjustment is excellent for the $B$-data set 
(without a primary minimum) and {fair} for the $V$-data set.
Weights were associated based on the overall standard deviations of the (CT~Her - C1) differential $B$-magnitudes computed 
night by night: while 1.0 was adopted in most cases, some sets with higher noise were allocated a relative weight 
of 0.2. This is the case for a few nights at the Beersel Hills Observatory (2005) and two nights 
at the Observatory of Sierra Nevada (2007). We restricted the residual data to the orbital phase bins between 0.05-0.95 
in order to {omit} the phase of primary minimum (sparsely covered by our V-observations). 
The fact that various outliers were found {close to a primary minimum} indicates that 
{ it is hard to model both effects simultaneously during this phase of rapid and steep light variation}. 
On the other hand,{ the outliers are only associated to} partially observed eclipses,
whereas the fully observed primary eclipse of 2008 shows normal residuals. Such small effects might also have been introduced by the airmass-dependent corrections (since more prominent at the beginning/end of the night). After the removal of a few outliers larger than or equal to 0.02 mag, the remaining standard 
deviations are {6.4 mmag in the $V$-band (with 1958 residual data points)} and {6.3 mmag in the $B$-band (with 7625 residual 
data points)}.\\

\subsection{Results from the $B$-data}\label{freqB}

We performed Fourier analyses with {\sc Period04} (\cite{lenz}). 
{Fig.~6 illustrates the frequency search in the range 0-80 \cd~ for the larger residual data set ($B$-band): the spectral window is shown in the top panel while the following panels show the initial periodogram and the periodograms successively prewhitened of the strongest signal from each previous run.
The last panel shows the residual periodogram with the mean noise level computed for frequency bins of width 5 \cd. We identified fifteen frequencies with an amplitude-to-noise ratio larger than or equal to 4.0 (the empirical criterion introduced by Breger et al.\ \cite{brg93}). The frequencies, amplitudes, residual standard deviations, signal-to-noise ratios and the removed fraction of the initial variance, $1 - R = (\sigma_{init}^2 - \sigma_{res}^2)/\sigma_{init}^2$, {obtained} from a multi-parameter fit of the residuals to a solution with fifteen frequencies, are listed in Table~\ref{results_freqB}. The errors on the frequencies and the amplitudes were computed using 120 Monte Carlo simulations of {\sc Period04}. The signal-to-noise ratios are based on the mean noise levels of the periodograms of the residuals. 
Remark that eight frequencies are concentrated in the range 43.5 -- 53.5 \cd. The main frequency of 52.93664 \cd~ has a semi-amplitude of 3.3 mmag in $B$. This frequency corresponds to a periodicity of 27.2 min, and is the one found by Kim et al.\ (\cite{kim}). A few oEA stars have{ even 
shorter main pulsation periods}: e.g. RZ~Cas ($f_1 = 64.1935$ \cd, Rodr\'{\i}guez et al.\ \cite{rod04}) and AS~Eri ($f_1 = 59.03116$ \cd, Mkrtichian et al.\ \cite{mkr2}).
\begin{figure*}[t]						    
\label{period04}							     
\resizebox{16cm}{!}{\includegraphics*{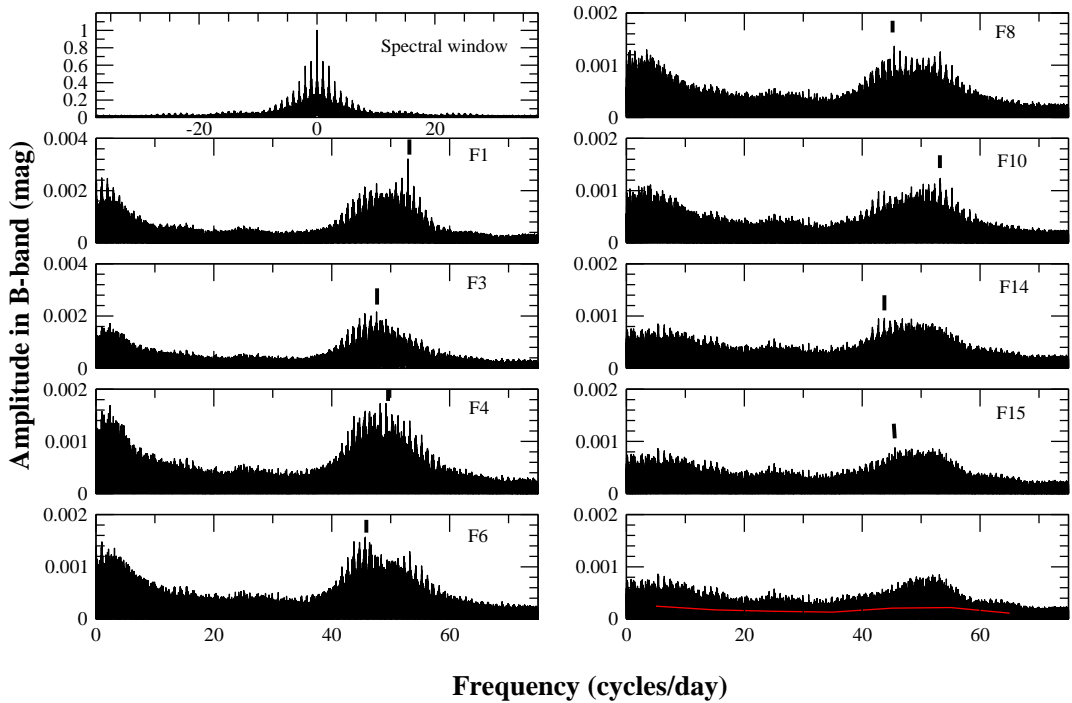}} 
\caption[]{Periodograms of CT~Her computed during successive frequency analyses (filter~$B$).} 					   
\end{figure*}		
					    


{All seven} frequencies in the range $f < 5$ \cd are caused by small imperfections linked to the reduction, extraction of the orbital variations and statistical fluctuations in the data sets. Since they are irrelevant for our study, we kept them fixed during the Monte Carlo simulations.
Some of the {other} eight frequencies found are affected by the 1 \cd~aliasing phenomenon: they are flagged 
in Table~\ref{results_freqB}. 
At this point, we conclude that eight (pulsation) frequencies are found to be significant. After prewhitening for the multi-frequency solution, the remaining standard deviation in the $B$-band equals 4.3 mmag, removing 53\% of the initial variance. High-quality residual $B$-band light curves 
are presented in Fig.~7,{ demonstrating a good} agreement between the residuals and the proposed solution.}\\ 


\begin{figure*}
  \begin{center}
    \begin{tabular}{ccc}
      \resizebox{50mm}{!}{\includegraphics{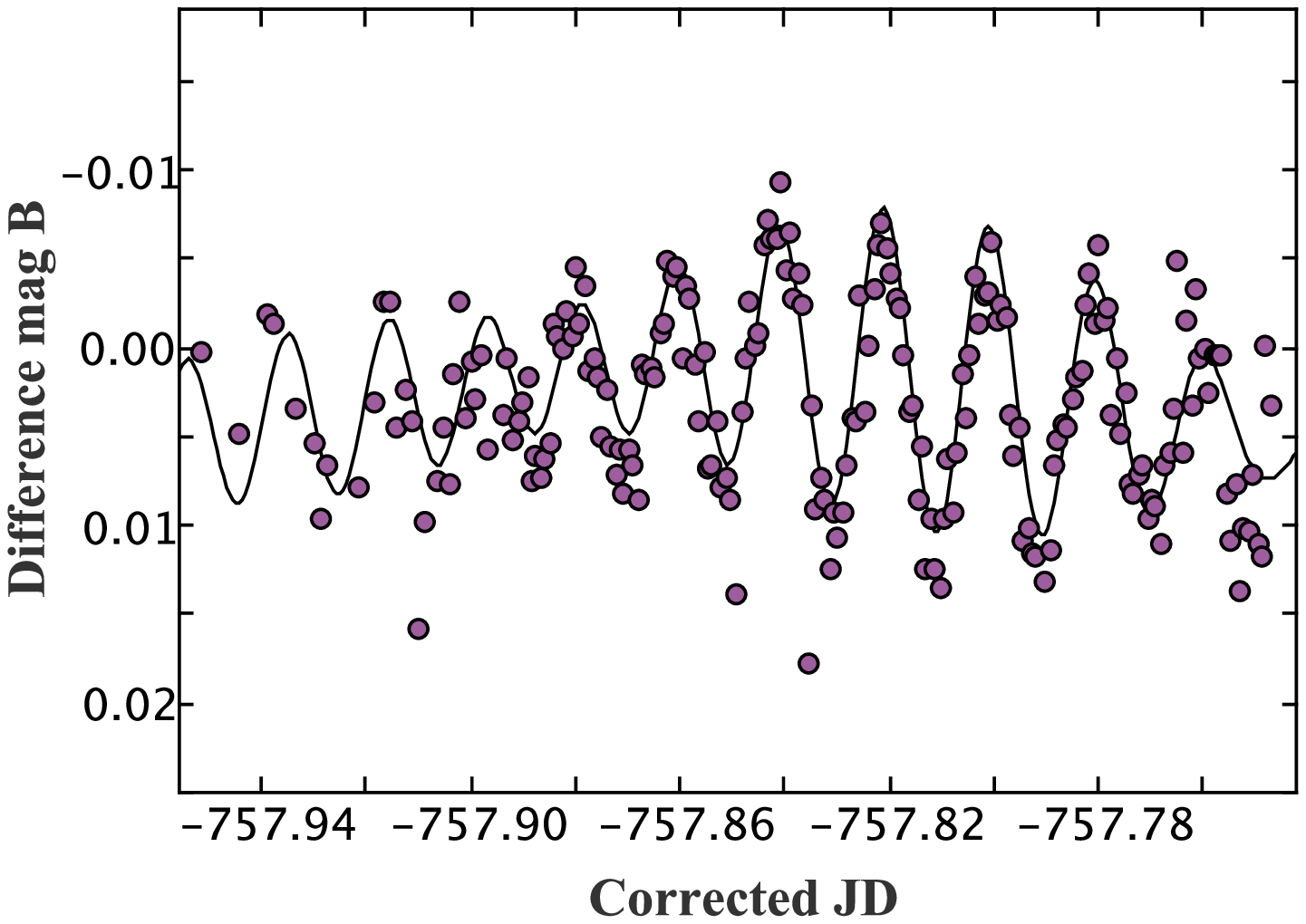}} &
      \resizebox{50mm}{!}{\includegraphics{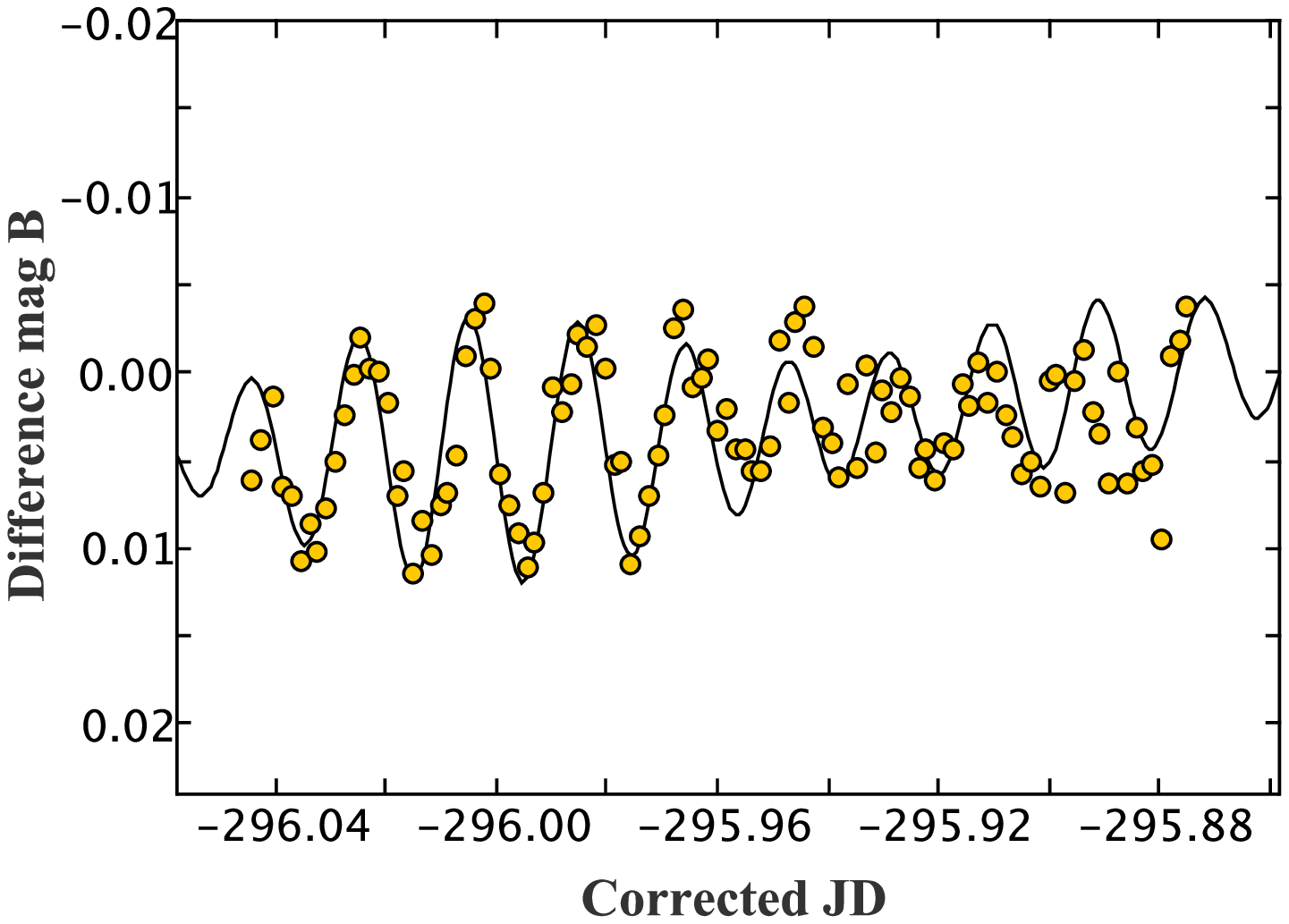}} &
      \resizebox{50mm}{!}{\includegraphics{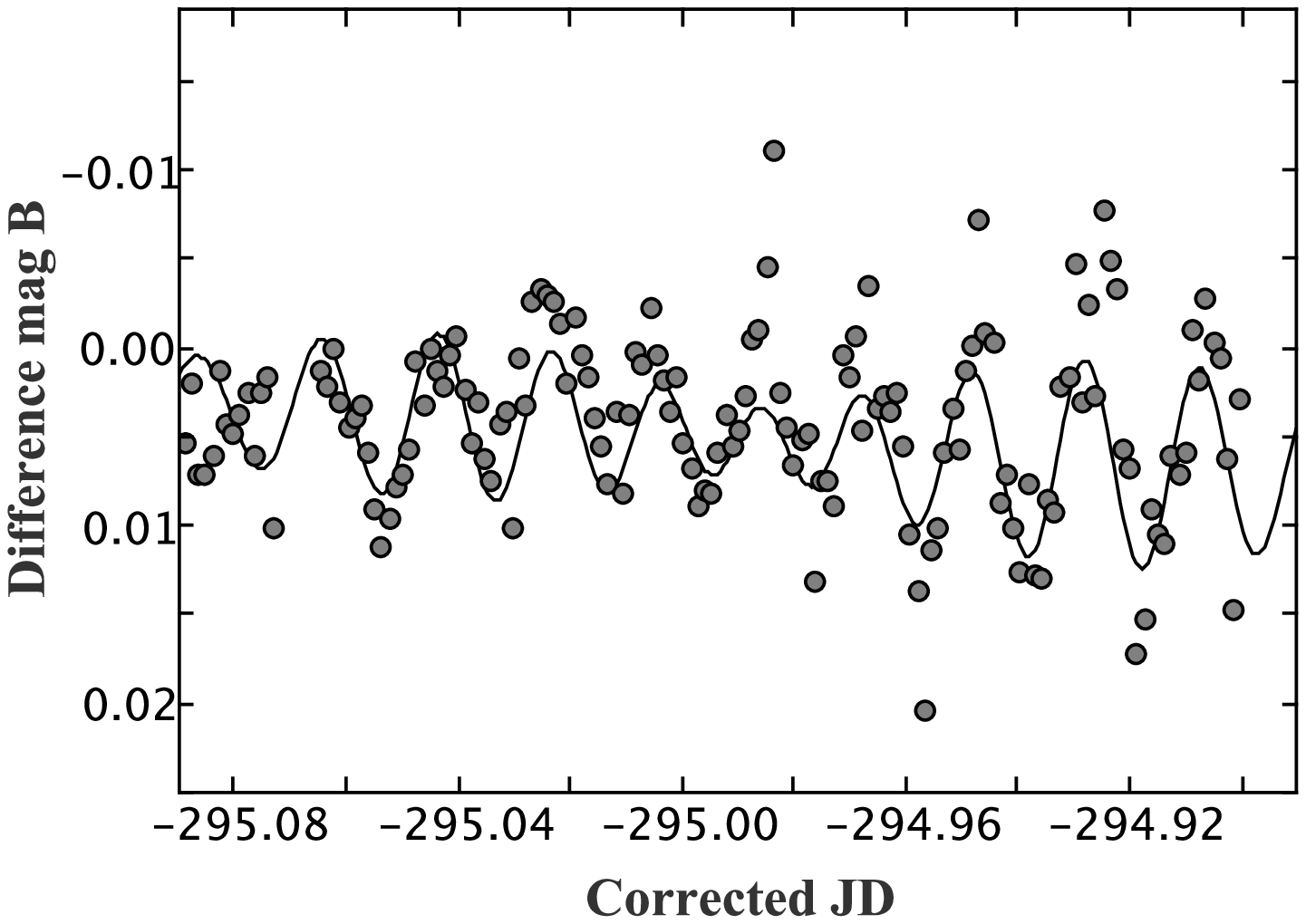}} \\
      \resizebox{50mm}{!}{\includegraphics{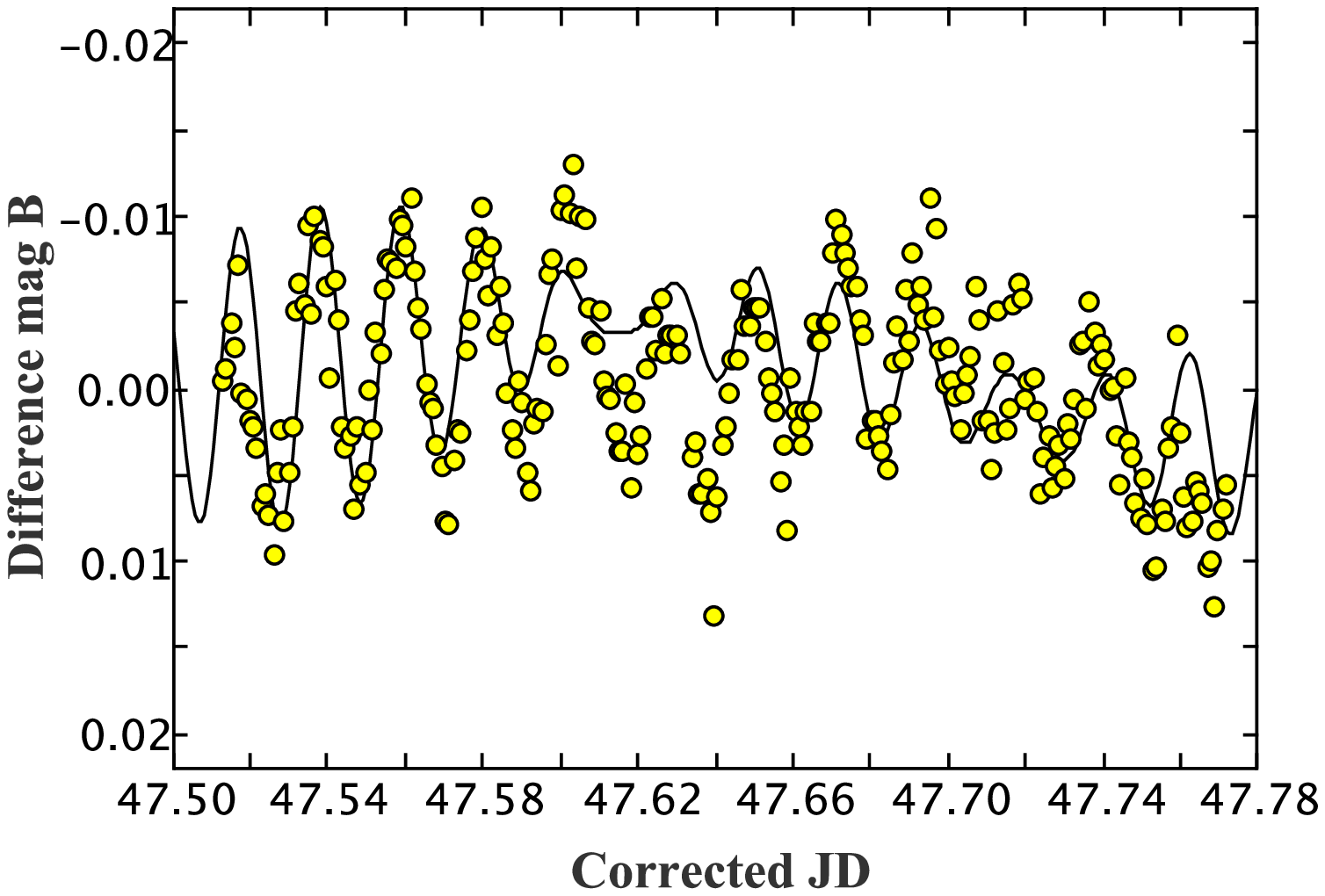}} &
      \resizebox{50mm}{!}{\includegraphics{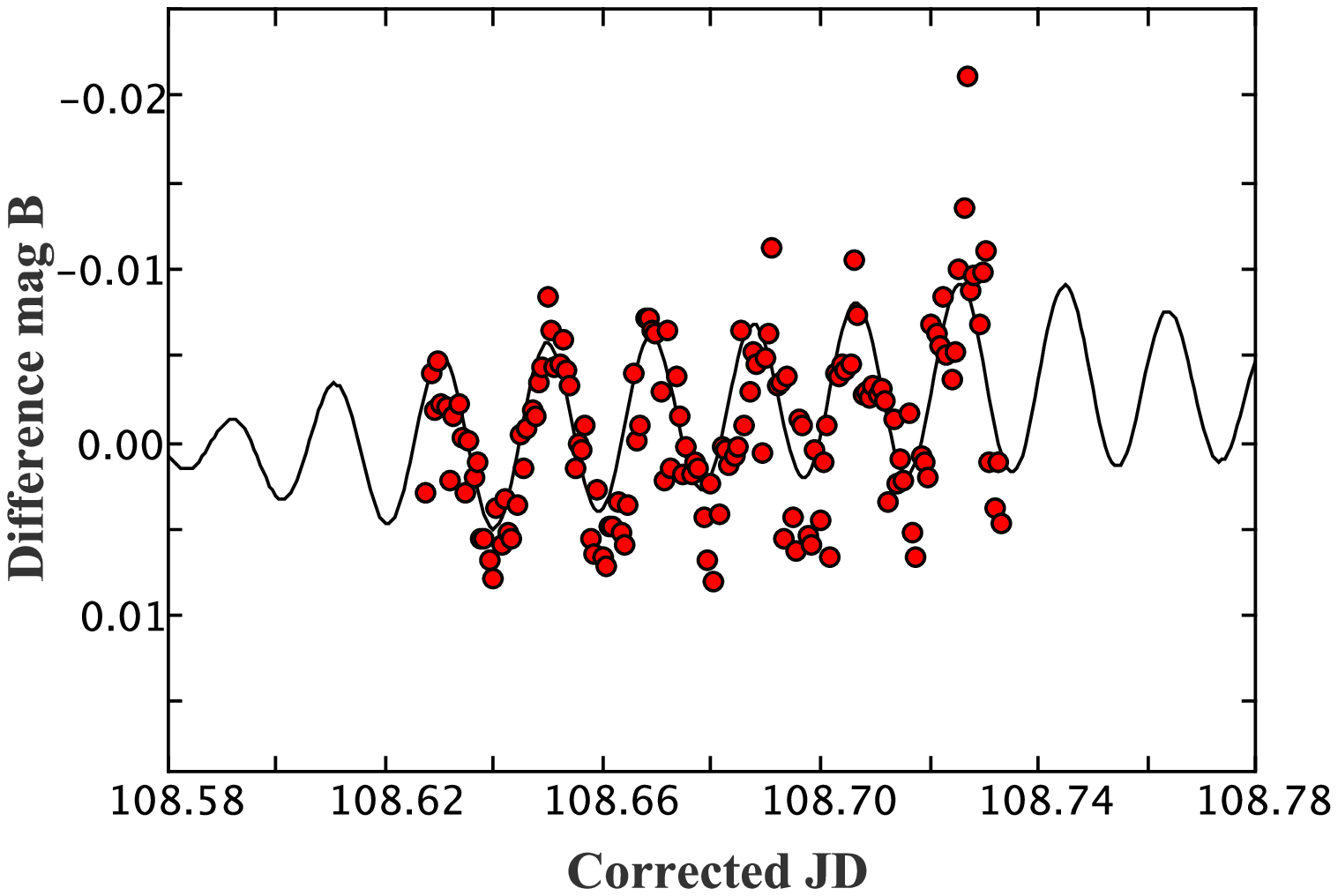}} &
      \resizebox{50mm}{!}{\includegraphics{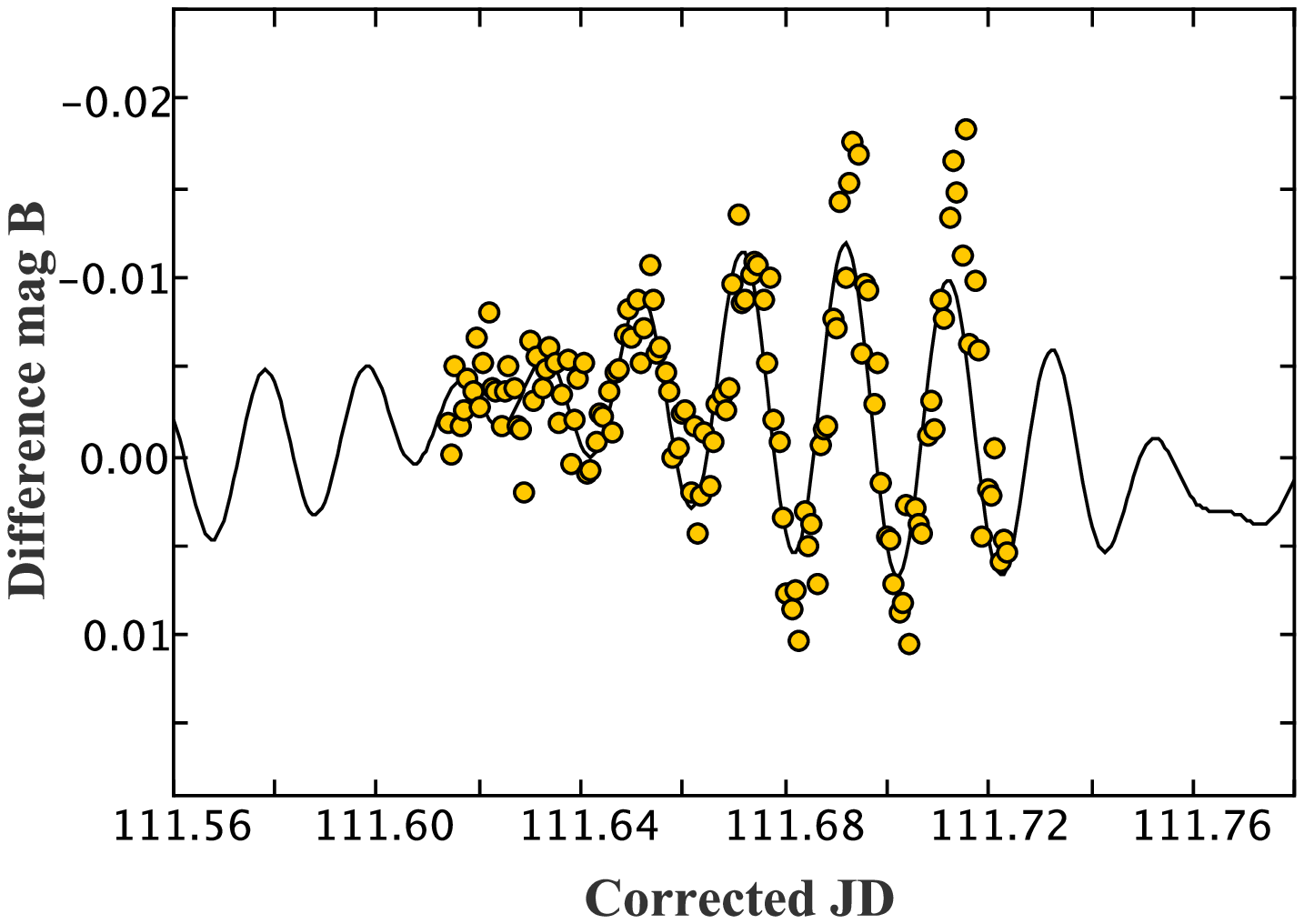}} \\
      \resizebox{50mm}{!}{\includegraphics{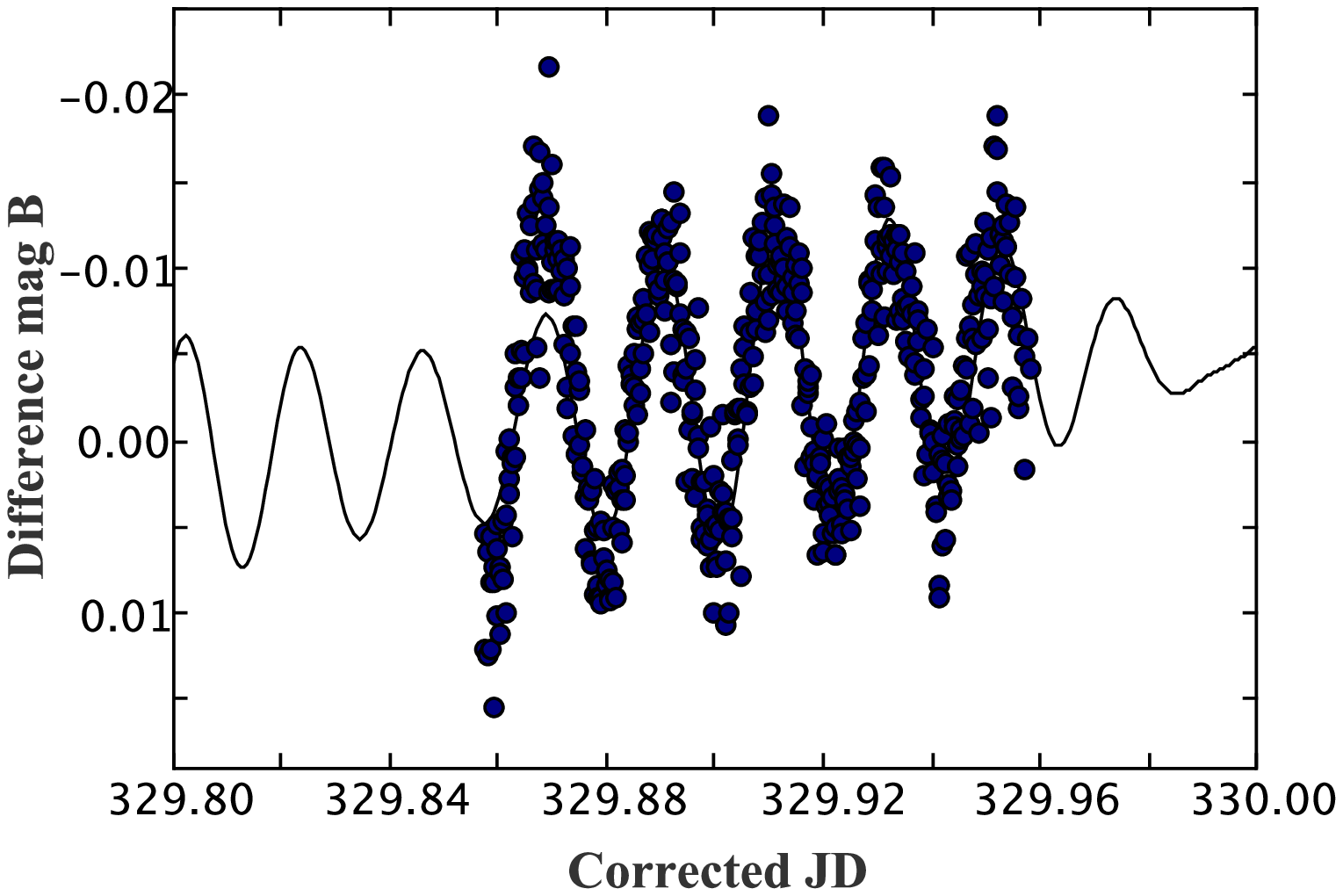}} &
      \resizebox{50mm}{!}{\includegraphics{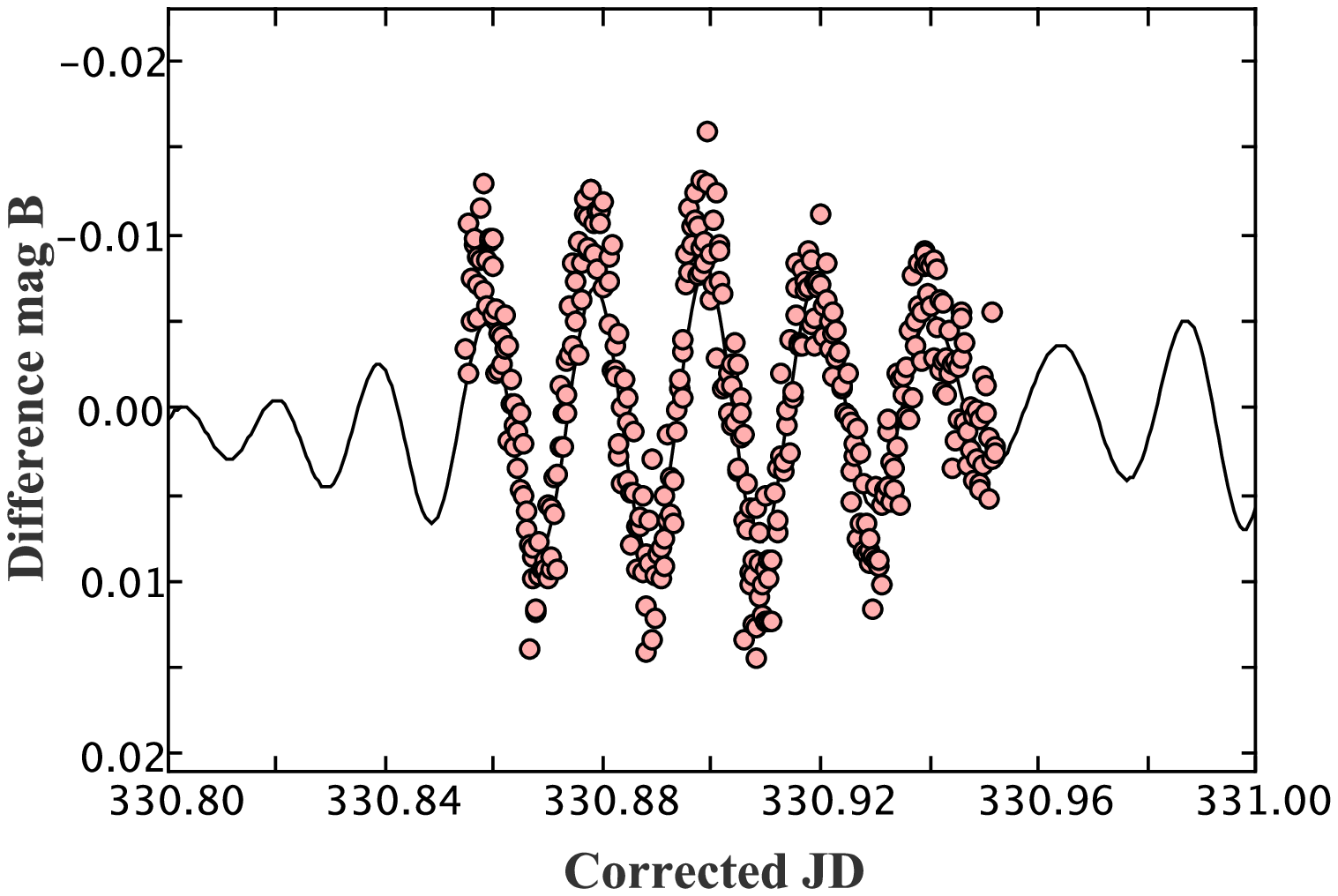}} &
      \resizebox{50mm}{!}{\includegraphics{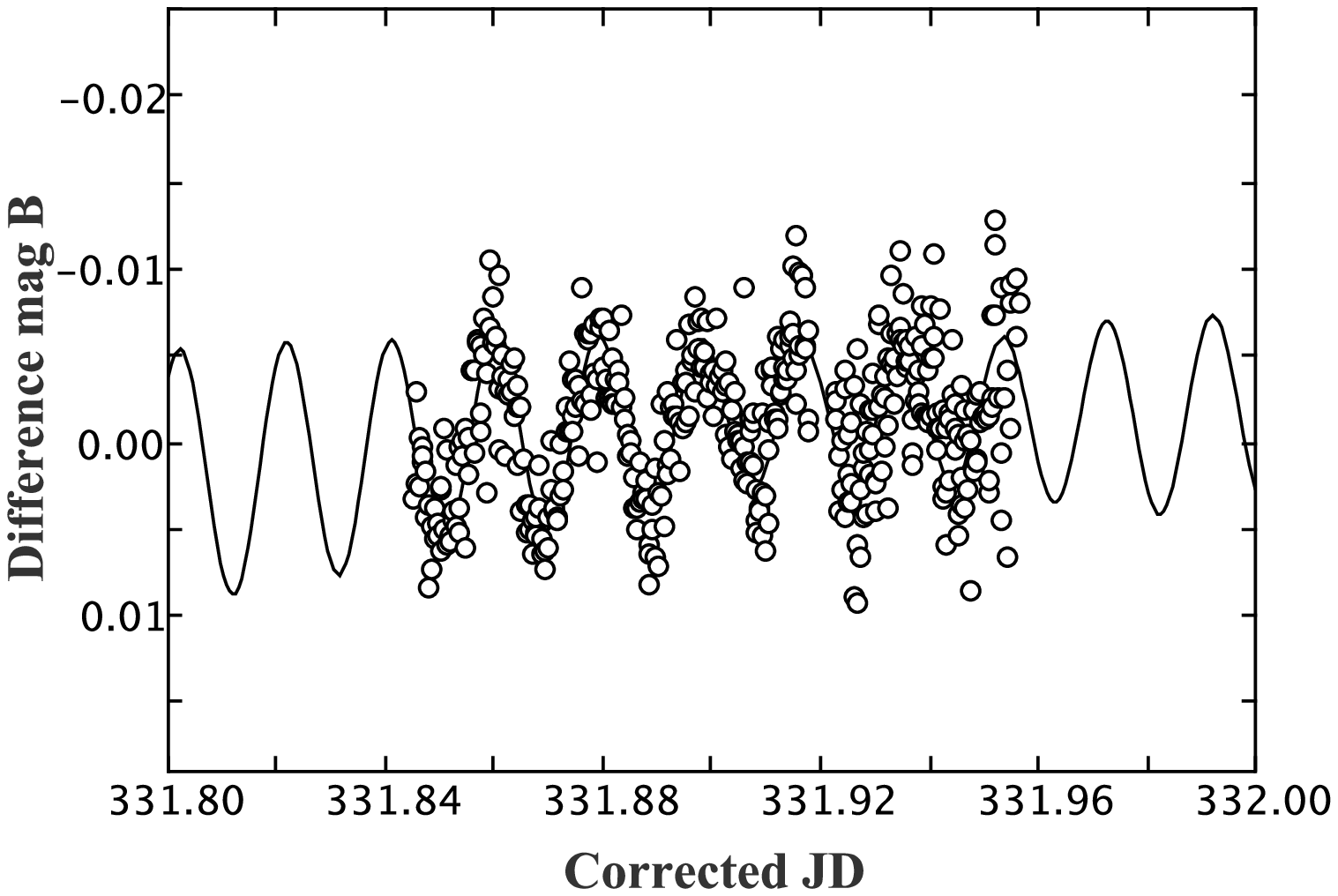}} \\
    \end{tabular}
    \caption{Sample figures of the $B$-band residual data set and corresponding multi-frequency model.}
    \label{9figs}
  \end{center}
\end{figure*}

\subsection{Results from the $V$-data}\label{freqV}

{ Likewise,} the complete set of the $V$-band residual data was analysed. Since these data are less numerous and more noisy, less significant results were found. We identified {ten} frequencies with an amplitude-to-noise ratio larger than 4.0, {six of which are located in the range 43.5 -- 53.5 \cd}. The frequencies, amplitudes, residual standard deviations, signal-to-noise ratios and the removed fraction of the initial variance, 1 - R, resulting from a multi-parameter fit of the $V$-band residual data set to a solution with{ ten} frequencies, are listed in Table~\ref{results_freqV}. All errors were computed using 120 Monte Carlo simulations of {\sc Period04}. Apart from {the frequencies satisfying $f < 5$ \cd, the frequencies G1 = F1, G3 = F4, G6 = F8 and G9 = F14 were recovered, thus confirming the presence of at least four of the eight}{ previously} detected frequencies. 
The {dominant} frequency has a semi-amplitude of 2.2 mmag and is located at 52.93762 \cd. It corresponds to the (1050 day)$^{-1}$ alias frequency of F1 due to a 3-yr gap in the time series. With respect to the $B$-band analysis, two unidentified frequencies, G4 (at 49.497 \cd) and G10 (at 48.068 \cd) were detected whereas the frequency F3 (at 47.600 \cd) was not{ confirmed}. Whether these new frequencies might possibly be related to those already found, remains unclear. Furthermore, the {frequencies F6 and F15}, which form an almost equally spaced triplet with G6 (= F8), were also not recovered. 
In Sect.~\ref{var_amp}, we will {discuss} these (apparently) partially incoherent results. After prewhitening for the solution with {ten} frequencies, the remaining standard deviation equals 5.1 mmag in the $V$-band. 

\begin{table}[t]
\caption[]{\label{results_freqB} $B$-band frequency-analysis of the residuals (model~72)}
\begin{tabular}{lccccc}
\hline 
ID  & Frequency    & Amplit. & $\sigma_{res}$  & S/N & 1 - R \struutup\\
    & ($\pm$ error) &  ($\pm$ 0.1)   &       &     &  \\
    & \cd        &   mmag    & mmag       &     &   \struutdown\\
\hline
F1    &  52.93664 ($\pm$ 0.8) & 3.3 & 5.9 &  12.7   & \struutup\\
F2    &   0.97682 (fixed)    & 2.8 & 5.6 &  11.4   & \\
F3    &  47.59996 ($\pm$ 2)  & 1.8 & 5.4 &   7.2   & \\
F4    &  49.20822 ($\pm$ 2)  & 1.8 & 5.3 &   6.6   & \\
F5    &   2.37738 (fixed)    & 2.0 & 5.1 &   8.1   & \\
F6    &  45.69130 ($\pm$ 3)* & 1.3 & 5.0 &  6.2   & \\
F7    &   1.00032 (fixed)    & 2.1 & 4.9 &   8.4   & \\
F8    &  45.44028 ($\pm$ 2)* & 1.4 & 4.8 &  6.8   & \\
F9    &   3.44103 (fixed)    & 1.1 & 4.7 &   4.3   & \\
F10   &  53.23747 ($\pm$ 2)  & 1.2 & 4.6 &   4.8   & \\
F11   &   4.08614 (fixed)    & 1.1 & 4.5 &   4.4   & \\
F12   &   0.61211 (fixed)    & 1.2 & 4.5 &   4.7   & \\
F13   &   4.16902 (fixed)    & 1.1 & 4.4 &   4.3   & \\
F14   &  43.76799 ($\pm$ 3)  & 1.1 & 4.4 &   5.4   & \\
F15   &  45.56395 ($\pm$ 4)* & 1.0 & 4.3 &  4.6   & 0.53 \struutdown\\
\hline
\end{tabular}

*: possibly affected by the 1 \cd aliasing
\end{table}

\begin{table}[t]
\caption[]{\label{results_freqV} $V$-band frequency-analysis of the residuals (model~72)}
\begin{tabular}{lccccc}
\hline 
ID  & Frequency    & Amplit. & $\sigma_{res}$  & S/N & 1 - R \struutup\\
    & ($\pm$ error) &  ($\pm$ 0.2)   &       &     &  \\
    & \cd        &  mmag    & mmag       &     &   \struutdown\\
\hline
G1    &  52.93762 ($\pm$ 5)$^{\&}$ & 2.2 & 6.4 &   7.4   & \struutup\\
G2    &   0.56716 (fixed)     & 2.2$^{e}$  & 6.1 &   6.4   & \\
G3    &  49.20636 ($\pm$ 6)$^{\#}$  & 1.8 & 5.9 &   6.8   & \\
G4    &  49.49661 ($\pm$ 7)   & 1.6 & 5.6 &   6.2   & \\
G5    &   3.68294 (fixed)     & 1.9 & 5.5 &   5.6   & \\
G6    &  45.43726 ($\pm$ 479)* & 1.1 & 5.4 &  5.3   & \\
G7    &   0.38478 (fixed)     & 2.3 & 5.3 &   6.8   & \\
G8    &   2.08342 (fixed)     & 1.6 & 5.2 &   4.7   & \\
G9    &  43.76795 ($\pm$ 22)  & 1.1 & 5.3 &   5.4   & \\
G10   &  48.06813 ($\pm$ 125)* & 1.1 & 5.2 &   4.4   & 0.36 \struutdown\\
\hline
\end{tabular}

*: (possibly) affected by the 1 \cd aliasing\\
$^{\&}$: is the 0.00095 \cd alias frequency of 5.93664 \cd (F1, Tab.~\ref{results_freqB})\\
$^{\#}$: is the 0.0018 \cd alias frequency of 49.20822 \cd (F4, Tab.~\ref{results_freqB})\\
$^{e}$: the error is $\pm$ 0.3 mmag\\
\end{table}



\subsection{Variability of the pulsation amplitudes}\label{var_amp}

The full analyses in the $B$ and $V$-bands were repeated after subtraction of the{ alternative} model using $T_1=8700$~K 
(model~71). {In the range above 24 \cd,} the frequencies listed in Tables~\ref{results_freqB} and~\ref{results_freqV} were {confirmed}, 
though with slight modifications in the order of their appearance. 
This shows that the results of the Fourier analyses are perfectly {\it independent} of the adopted choice 
for the binary model ({ owing} to a) the fact that the binary models are {almost} identical and b) the different time 
scales involved). 

It would {also} seem that{ the results of the frequency analyses in both filters} tend to a common solution, were it 
not for the presence of F3 (Table~\ref{results_freqB}) and G4 (Table~\ref{results_freqV}). In addition, the order in which the frequencies 
were detected in the respective data sets is not identical. Therefore, we investigated whether patterns could be found in the pulsation amplitudes 
assuming that the frequency content is stable. All eight frequencies from Table~\ref{results_freqB} located in the range 43.5 -- 53.5 \cd~
were adopted to represent the frequency content of CT~Her. Preference was given to the results from the $B$-band analysis because the aliasing is much stronger in the $V$-band data (particularly the 1 \cd~and the 0.0018 \cd~aliasing effects).
Next, we computed the multi-parameter solutions with the amplitudes and the phases as free parameters on a yearly basis.

\begin{table}[]
\caption[]{\label{amp_var} Semi-amplitudes of the pulsation frequenties ($B$-band)}
\begin{tabular}{ccccc}
\hline 
 ID &  Amp.   &   Amp.  &  Amp.  & Amp. \struutup\\ 
    &   2004     &   2005   &  2006   & 2007-2008 \\ 
    &  {\tiny ($\pm$ 0.3)}   & {\tiny ($\pm$ 0.2)}  & {\tiny ($\pm$ 0.1)} & {\tiny ($\pm$ 0.1)} \\ 
    &    mmag         & mmag         &  mmag       &  mmag \struutdown\\
\hline
No. data       &  951  &  1704  &  2262   &   2511 \struutup\\ 
$\Delta$T     &   97  &    61  &    72   &    466 \struutdown\\ 
\hline
F1 &  3.0  & 3.2 & 3.4 & 3.2 \struutup\\ 
F3 &  2.2 {\tiny ($\pm$ 0.4)} & 1.1 &  2.2$^{e}$ & 1.6$^{e}$ \\
F4 &  2.4 & 2.1 &  1.3  & 2.0$^{e}$ \\
F6 &  2.2 & 1.7 &  0.7$^{e}$ & 1.7\\
F8 &  0.9 & 0.6 &  1.4$^{e}$ & 2.2\\
F10 &  1.8 & 1.2 &  1.2  & 1.8\\
F14 &  0.2 & 1.0 &  1.5  & 0.8\\
F15 &  1.3 {\tiny ($\pm$ 0.5)} & 0.5 {\tiny ($\pm$ 0.5)} &  1.2$^{e}$ & 1.2$^{e}$ \struutdown\\
\hline
\end{tabular}

$^{e}$: the error is $\pm$ 0.2 mmag\\
\end{table}

The results of{ the computations} are shown in Table~\ref{amp_var} and illustrated in Fig.~10.
{While the amplitude of the frequency F1 can be considered to be stable over the entire observing season, this is not true for all the frequencies. 
The frequencies F4 and F6 behave similarly (first decreasing in amplitude until 2006 and then increasing), whereas the frequencies F3 and F8 show a maximum of their amplitude, respectively in 2006 and 2007. In contrast, 
the frequencies F14 and F6 show variability of their amplitude which occurs in anti-phase. This behaviour explains why the frequencies 
detected 
in Sect.~\ref{freqB} and~\ref{freqV} are not identical: the detection of G1 = F1, G3 = F4 and G4 = F8{ may} be understood in the light 
of the 2005 (partially supplemented by the 2007-2008) $B$-band semi-amplitudes. 
In conclusion, the amplitudes of these pulsation frequencies show evidence of variability on time scales of 1-2 years, perhaps even less, with the exception of the most dominant frequency which has a stable amplitude.}

\begin{figure}[]
\centering
\label{Fig7}
\resizebox{6.5cm}{!}{\includegraphics*{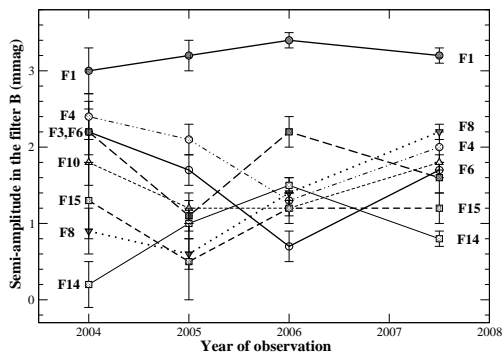}}           
\caption[]{Semi-amplitudes of the pulsation frequencies versus time (filter $B$)}
\end{figure}

\section{Radial velocities}\label{spectro}

\begin{figure}
\hspace*{0.5cm}
\centering
\resizebox{8cm}{!}{\includegraphics*{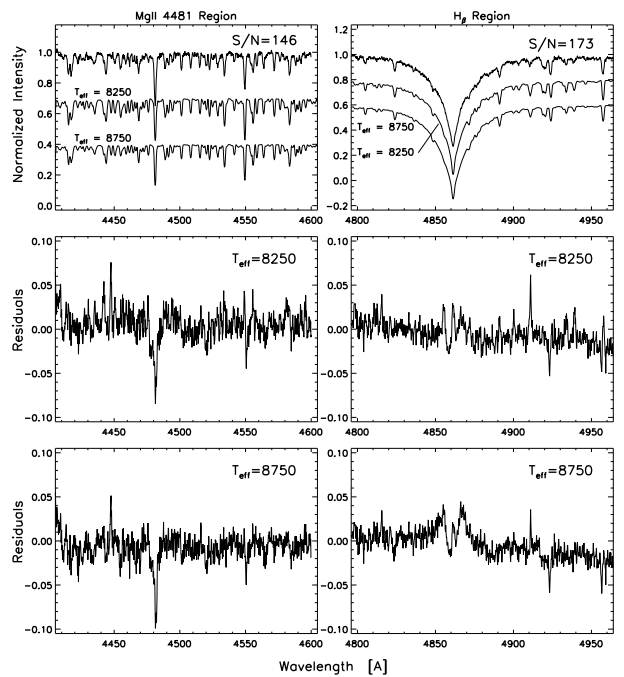}}           
\caption[]{Top panel: combined observational spectra and synthetic spectra showing model~A ($T_{\rm eff} = 8250$~K), and model~B ($T_{\rm eff} = 8750$~K). Other panels: difference spectra in the sense observed minus model~A (middle) and observed minus model~B (bottom).}
\label{spectrumpart}
\end{figure}

{ High-resolution spectra of CT~Her} were gathered with the 2-m~telescope of the NAO, Rozhen, and the Coud\'e spectrograph equipped with an AT200 Photometrics camera (
with a SITe~SI003AB chip and pixel size of 24 $\mu$m), and a spectral resolution of 0.19 \AA/pixel. {Three regions were observed}: H$_{\beta}$ and the regions around MgII ($\lambda = 4481\AA$) and FeI ($\lambda = 5455\AA$). Forty-seven spectra were collected during 17 nights: 5 nights in May-July 2007, 5 nights in April-June 2008, as well as 
7 nights in April-July 2009. Typical exposure times were 1200 and 1800 s. The average signal-to-noise ratio S/N is $\approx$ 30. The spectra were reduced with standard IRAF procedures. The corresponding radial velocities were measured with the cross-correlation technique using synthetic spectra obtained with the programme SPECTRUM (\cite{gr}), and a grid of LTE-atmosphere models of solar-type chemical composition (\cite{ca})(Table~\ref{radvel}). Combined spectra having S/N $>$ 100 were used to estimate the effective temperature of CT~Her~A: the best fit for two regions (H$_{\beta}$, MgII) was obtained with $T_{\rm eff} = 8250$~K, $\log g = 3.7$ and $v\sin i = 60$ km/s (Fig.~\ref{spectrumpart}).

\begin{table}[t]
\caption[]{\label{radvel} Radial velocities of CT~Her~A (first three lines, table available in electronic form) }
\begin{tabular}{lcc}
\hline
HJD           &  RV    & Error\\ 
\hline                                                                                                
2454247.47535 & -44.21  & 3.20\\
2454247.49155 & -52.88  & 2.21\\
2454248.37028 & -1.60   & 2.63\\
\dots         &           &   \\
\hline
\end{tabular}
\end{table}

\begin{figure}[t]
\centering
\resizebox{8cm}{!}{\includegraphics*{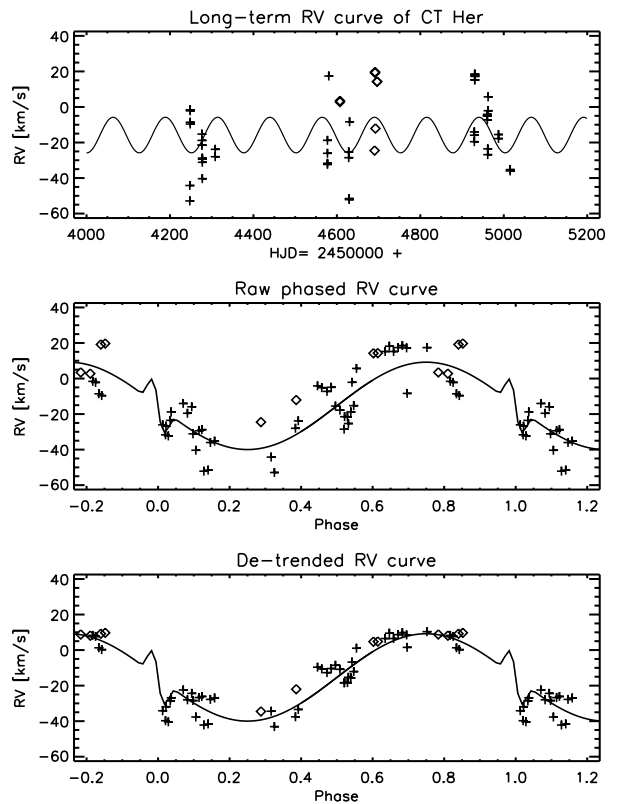}}           
\caption[]{Radial velocities of CT~Her~A obtained from NAO spectra (plusses) and CLA spectra (diamonds): a 125.3 days
{periodicity} was removed before fitting in combination with the $B$- and $V$-light curves.} 
\label{RaVel}
\end{figure}

Eight spectra were also collected with the {\sc Foces} spectrograph attached to the DSAZ 2.2-m telescope of the Calar Alto Observatory (CLA) in 2009. The resolution is 
40600. The spectral region{ covers} 5000-6000 \AA. The average S/N is $\approx$ 25.  The radial velocities of both components were measured with
the cross-correlation technique using synthetic spectra generated for temperatures of 8200~K with $\log~g~=~4.0$ and 4500~K with $\log~g~=~3.5$, and for $v \sin i~=~50$ km/s. In this case, the 2-D cross-correlation {programme} {\sc Todcor} 
was used (\cite{maz}). All radial velocity measurements are presented in Table~\ref{radvel}, available in electronic form only. We first checked that the radial velocities obtained for component~A were compatible with the NAO radial velocities. In a plot versus time (cf. top panel of Fig.~\ref{RaVel}), we detected a long-term variation which had to be removed{ before} fitting. 

A Fourier analysis of the residuals obtained from the raw velocities performed with {\sc Period04}, and a preliminary fit with velocity 
semi-amplitude $K_A = 23.4$ km/s and systemic velocity $\gamma = -15.8$ km/s indicated a periodicity of $\approx125.3\pm2.2$ days with semi-amplitude $K_C = 11.4$ km/s. The corresponding ephemeris can be described by: $HJD_{max}~=~2454157.6~+~125.3\times E$.
Fig.~\ref{RaVel} shows the model (solid line in top panel) and the {raw and corrected} radial velocity curves (bottom panels). The standard deviations of the radial velocity residuals amount to 11.69 km/s and 7.63 km/s, respectively for the {raw} data and for the data after subtraction of the model.

\begin{figure}[]
\centering
\resizebox{7cm}{!}{\includegraphics*{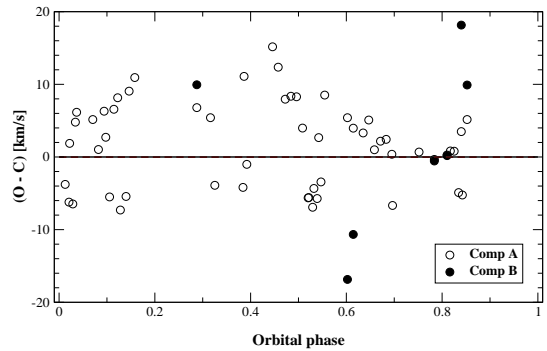}}         
\caption[]{Residual component radial velocities of CT~Her.}
\label{resRV12}
\end{figure}

The {corrected} radial velocity curves were next used in combination with the $B$ and $V$ light curves to enable fitting of the following parameters with {\sc Phoebe}: the semi-axis major expressed in solar radii, A, and the systemic velocity, $\gamma$. The new parameter values and their formal uncertainties are listed in Table~\ref{model}. This value of {the semi-axis major}, A, lies close to its approximate value (\cite{sv}). The values of the $\chi{^2}$ function are slightly worse than before. This new solution fits both light curves a little {worse}, particularly the primary eclipse in the V-filter, but accomodates the radial velocities reasonably: the means with their standard deviation of the residual radial 
velocities equals $1.7\pm5.8$ km/s and $1.4\pm12.3$ km/s, for component~A and B respectively (Fig.~\ref{resRV12}). The main differences with the former solutions are the smaller value of the mass ratio {($q=0.127$)} and the higher value of the inclination {($i = 82.8\degr$)}. 
Table~\ref{phys} summarizes the physical properties of both components resulting from the combined data fitting. These properties appear to be compatible with the status of component~A as a pulsator of type $\delta$ Scuti. 
The computed masses are {furthermore consistent} with those preliminarily derived by \cite{hoff} from spectroscopy only (they obtained ${\rm M}_1 = 2.31\pm0.11$~\Msol~and ${\rm M}_2 = 0.31\pm0.09$~\Msol, with $q = 0.13$ (in agreement with the q-values of Tables~\ref{t71} and~\ref{model})).
We propose that this new solution illustrates the uncertainty still remaining in the{ characterization} of the physical parameters of CT~Her: {more realistic errors would thus correspond to the formal errors of Table~\ref{t71} (Table~\ref{model}) multiplied with a factor of 25-50 while an uncertainty of about 80~K would be closer to the truth in the case of $T_{2}$}. The data collected at the phase of primary minimum are {probably} affected by the pulsations (this is indeed the case for other oEA stars), which may introduce some indetermination in the light curve modelling. However, our radial velocity data, in particular of component B, are too scarce 
to enable an accurate determination of the mass ratio. Accurate component radial velocities obtained from high-resolution \'echelle spectra 
would be necessary to obtain a {more} consistent determination of the absolute parameters of CT~Her. 
Meanwhile, the solution(s) derived in Sect.~\ref{LCmodel} are currently the most adequate one(s).

\begin{table}
\caption{Parameters of the combined light and radial velocity solution for CT~Her.}
\label{model}
\begin{tabular}{lcc}
\hline
\mcol{1}{l}{Parameter}  & \mcol{1}{c}{Filter $B$ } & \mcol{1}{c}{\hspace{1.5mm} Filter $V$ } \struut\\
\hline
$i$ $(\degr)$       & \mcol{2}{c}{82.75     $\pm$ 0.04   } \struutup\\
$q$                 & \mcol{2}{c}{0.1267    $\pm$ 0.0006 } \\
$A$                 & \mcol{2}{c}{8.48      $\pm$ 0.03   } \\
$\gamma$ $(km/s)$   & \mcol{2}{c}{-16.2     $\pm$ 0.2    } \\
$T_{1}$ (K)         & \mcol{2}{c}{8200$^{a}$             } \\   
$T_{2}$ (K)         & \mcol{2}{c}{4468      $\pm$ 7     } \\
$\Omega_{1}$        & \mcol{2}{c}{4.269     $\pm$ 0.008  }  \\
$\Omega_{2}$ $^{c}$ & \mcol{2}{c}{2.038                  } \\ 
$[{L_{1} \slash (L_{1} + L_{2})}]$ & 0.969 & 0.932 \\
$[{L_{2} \slash (L_{1} + L_{2})}]$ & 0.031 & 0.068 \\
$g_1$               & \mcol{2}{c}{1.0$^a$      }  \\
$g_2$               & \mcol{2}{c}{0.32$^a$     }  \\
$A_1$               & \mcol{2}{c}{1.0$^a$      }  \\
$A_2$               & \mcol{2}{c}{0.5$^a$      }  \\
$x_{1}$ $^{c}$      & 0.585$^b$         & 0.505$^b$ \\
$x_{2}$ $^{c}$      & 0.962$^b$         & 0.811$^b$ \struutdown\\
\hline
No. data     & 7960   & 2180   \struutup\\
$\chi^2$    & 0.0162 & 0.0055 \\
No. RV data  & 55 (comp~A) & 7 (comp~B)\\
$\chi^2$    & 0.0038 & 0.0151 \struutdown\\
\hline
\end{tabular}

$^{a}$ \small {adopted value} \\
$^{b}$ \small {Van Hamme (1993) tables} \\
$^{c}$ $x_{1}$, $x_{2}$ are the limb darkening coefficients. $\Omega_2$ is the dimensionless potential of the secondary component.\\ 
\end{table}

\begin{table}
\caption{Tentative absolute parameters of CT~Her}
\label{phys}
\centering 
\begin{tabular}{lcc}
\hline
\multicolumn{1}{c}{Parameter} & \multicolumn{1}{c}{Comp A} & \multicolumn{1}{c}{Comp B}\\
\hline
Mass (\Msol) &$2.28 \pm 0.01$ & $0.29 \pm 0.04$ \\
Radius (\Rsol) & $2.06 \pm 0.06$ & $1.87 \pm 0.08$ \\
$T_{\rm eff} (K)$  &  $8200$ & $4468 \pm 80$\\
log~g & 4.17 $\pm$ 0.02 & 3.36 $\pm$ 0.02\\
Luminosity ($\rm{L_\odot}$) & $17.4 \pm 2.4$ & $1.2 \pm 0.2$ \\
$M_{\rm bol}$ (mag) & $1.65 \pm 0.15$ & $4.50 \pm 0.21$  \\
\hline
\end{tabular} 
\end{table}

\section{Possible effect on the pulsation analysis}\label{LTT_phases}

\begin{figure}[]
\centering
\resizebox{7cm}{!}{\includegraphics*{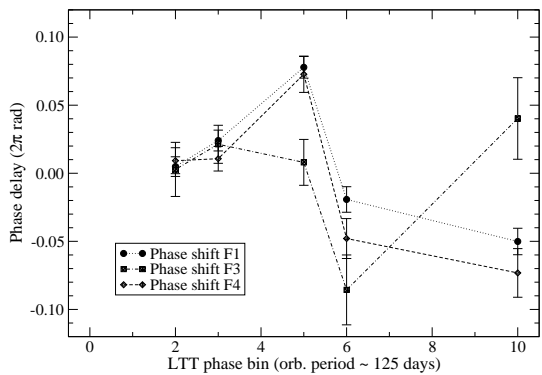}}         
\caption[]{Phase shifts computed for the pulsation frequencies F1, F3 and F4 as a function of the orbital 
phase bin (with respect to the long periodicity).}
\label{LTT_phaseshift}
\end{figure}

{ The presence of a third body can cause a cyclic variation of the systemic velocity. In that case, 
the resulting} light travel time (LTT) effect will introduce periodical{ time delays} in 
the pulsational analysis. We examined if {any} phase shifts could be detected by subdividing the $B$-filter data 
into 10 subsets arranged according to the orbital phase bin associated with the long-term ephemeris 
discussed in Sect~\ref{spectro}. We used the multivariate analysis method implemented 
in {\sc PERIOD04} (\cite{brg05}) to recompute the best-fitting phases 
for each subset with respect to the proposed multi-frequency solution. The outcome of these 
computations is that we find no obvious shift with respect to the initially adopted phases. 
Fig.~\ref{LTT_phaseshift} represents the computed phase shifts, and illustrates that there is 
no detection of a { systematical} periodical pattern in the phases of the frequencies F1, F3 and F4. 
It is clear that a lot of data in each subset are necessary to be able to reliably use the result of the 
multi-frequency fit. That is why we considered 6 out of 10 subsets only. 
For these frequencies (F1, F3 and F4) and for these subsets, the errors on the phases are sufficiently precise.   

{Alternatively, one can 
derive a model making allowance for a periodic variation of the phase shifts of the multi-frequency solution. 
This approach has been applied as well and the results 
confirm that no significant differences on the (pulsation) frequencies are detected even though a slightly better overall 
fit was obtained, corresponding to a 129.6~days LTT (solution with $\sigma_{res} = 4.2$ mmag). 
}

\section{The O-C diagram}\label{ominc}

\begin{table}
\caption[]{\label{octab} O-C values for two models representing the evolution of P$_{orb}$ (first three used lines, table available in electronic form) }
\begin{tabular}{lccccc}
\hline
HJD            &  E & O-C$_{1}$ & O-C$_{2}$ & p/s & Tech.\\ 
\hline                                                                                                
\dots          &&&&&      \\
2438894.7540 & -2031 & -0.0508 & -0.0337 & p & vis\\
2438953.7280 & -1998 & -0.0271 & -0.0103 & p & vis\\
2439248.5090 & -1833 &  0.0020 &  0.0175 & p & vis\\
\dots          &&&&&      \\
\hline
\end{tabular}
\end{table}

\begin{figure}[]
\centering
\resizebox{7cm}{!}{\includegraphics*{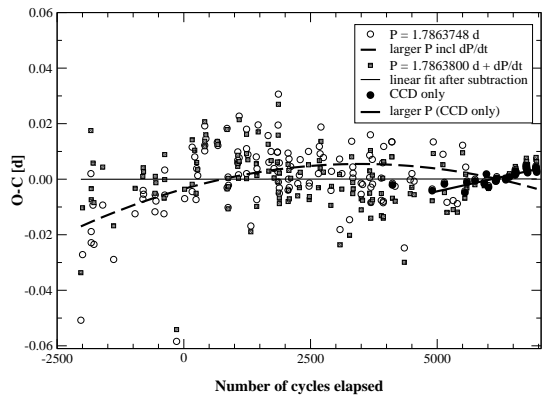}}           
\caption[]{Evolution of the orbital period: O-C diagrams of CT~Her using a constant and a variable model.}
\label{ominc3}
\end{figure}

The {same} LTT effect also implies a cyclic variation of the orbital period. Therefore, we investigated the O-C diagram of CT~Her. We used the list 
of observed times of light minima from the database of the Variable Star and Exoplanet Section of the Czech Astronomical Society (\cite{oc}), 
from which we selected 146 observed times collected between 1965 and 2009 ($JD > 2438761$), after elimination of the 26 first epochs. Then,
we added two of our own observed minima (one {minimum} was not useful for the modelling of the $B$-band light curve because of an unexplained 
sudden drop during the eclipse and another isolated minimum was obtained by D. Litvinenko in an unreported filter): $2453189.3784\pm0.0002$ ($E=5971$)
and $2453514.4961\pm0.0001$ ($E=6153$). Using $P_{orb} = 1.7863748$ days (\cite{sam}), we recomputed the O-C values and searched for 
the best representation using either a linear or a parabolic model. Fig.~\ref{ominc3} illustrates{ both} cases: the circles show the O-C values 
with the known orbital period while the squares show the O-C values using the ephemeris corresponding to the best-fit parabola:
\[ HJD_{min} = 2442522.92914 + E\times1\fd7863799 \] \[\hspace{1.5cm} - E^2 \times 7\fd394 \times10^{-10}.\] 
Such a model implies a larger orbital period with a decrease dP/dt of $8.28\times10^{-10}$ days~\cd (i.e. 
a secular period decrease dP/(Pdt) of $169\times10^{-9}{\rm yr}^{-1}$). {The rms is 0.01049~days (906~s). 
To check its significance {with respect to the linear model}, we used an upper one-tailed F-test to verify whether the null 
hypothesis that the sums of squared residuals of both models are equal holds. 
The test statistic equals
\[((\chi_{lin}^2 - \chi_{par}^2) \times (n-3)) / (\chi_{par}^2 \times (3-2)) = 23.44.\]
Since this value is much larger than the tabulated value $F_{0.01}(1,145) = 6.81$ (\cite{pez11}), 
we reject the null hypothesis at the 1\% significance level (type~I error), meaning that the parabolic model is statistically
significant. We remark that \cite{soy08} used all previous O-C data (including the 26 first measurements) to derive an 
overall increase with a cyclic variation for the orbital period of CT~Her. For evident reasons, we prefer to discuss only the 
well-observed part of this diagram.} 

Table~\ref{octab} lists the cycle numbers and the associated residual values for both models. However, the first reported times do not 
fit these models at all as is also obvious from the O-C diagram in the {\it Atlas of O-C diagrams of Eclipsing Binary Stars} (\cite{kre}). 
Moreover, the scatter remains large in both cases. Interestingly, if we restrict the observations only to the 24 most recent times observed 
using the CCD technique, we obtain a tight-fitting linear model with a high correlation coefficient (0.9) indicating an increased orbital 
period of 1.7863789 days with respect to the known period (cf. the speckled circles in Fig.~\ref{ominc3}). The rms in this case is 
only 0.00135~days (117~s). 
 
Using the information about a probable variation in systemic velocity with a period of 125.3 days and a semi-amplitude of 11.4 km/s 
(Sect.~\ref{spectro}), we estimate a mass function equal to
\[f(M) = 1.0385 \times 10^{-7} \times (1 - e^{2})^{3/2} \times K_{C}^{3} P = 0.019~{\rm M_{\odot}}\] 
and a corresponding O-C amplitude of 131~s. Because this is of the same order as the computed rms 
(117~s), we conclude that the LTT effect is barely observable from the currently available data and that the O-C diagram of CT~Her does not 
contradict the proposed model of a variable systemic velocity due to the presence of a third body. {Remark that the first two
epochs} of light minimum observed with a CCD show a flat trend, in full agreement with the period mentioned in the literature. 

\section{Summary and conclusions}

{ CT~Her is yet another oEA star}, after Y~Cam, in which a large set of pulsation frequencies has been identified thanks to a long-term photometric monitoring {performed} in two filters 
{ We confirm the presence of rapid pulsations with a dominant frequency of 52.93664 \cd.} 
{As many as eight} significant frequencies, all of which are located in the range 43.5-53.5 \cd,
appear to reproduce the observed, {complex} behaviour reasonably well throughout the years 2004-2008. 

Furthermore, {except for the most dominant frequency,} we showed that the amplitudes of these pulsation frequencies are variable on time scales of 1-2 years, perhaps even less than one year, assuming that the frequency content is stable (the fact that several frequencies are commonly found in different data sets suggests this). Considering the absolute parameters of the (actual) primary component of CT~Her (Table~\ref{phys}), we conclude that the dominant frequency 
cannot correspond to the fundamental radial mode: {rather,} such a short period of pulsation for a $2.5$~\Msol~star {is} typical of a high overtone radial mode or a non-radial pulsation mode of high order ''p'' (similar to RZ~Cas, AS~Eri or TW~Dra). Using the {improved} ephemeris, we {derived} an accurate value of 94.56 for the ratio $P_{\rm orb}/P_{\rm puls}$.

{ The analysis of} recent spectra providing{ complementary} radial velocities suggests that the light and radial velocity curves 
can be{ modelled simultaneously}, provided that the radial velocity measurements {are corrected for} a long periodicity of $\approx$ 125 days with a semi-amplitude $K_C = 11.4$ km/s. The presence of a third body in the system could explain such a light travel time effect. However, we were not able to confirm this suggestion from an investigation of the O-C diagram of CT~Her as the expected O-C amplitude is of the {same} order {as that of} the noise. {Neither did we detect an {obvious} orbital modulation in the phases of the multi-frequency solution.}

Although the match between the model and the $B$- and $V$-light curves is excellent,{ we cannot yet claim to know} the system parameters of 
CT~Her uniquely,
especially because accurate component radial velocities are lacking. {Such component radial velocities are much needed} in order to obtain a consistent determination of the absolute parameters of this Algol-type binary. {Meanwhile, we acquired a new series of high-resolution spectra of CT~Her with the Hermes 
spectrograph attached to the Mercator telescope (\cite{raskin}). High signal-to-noise spectra {should} also be very useful for application of} the technique of spectra disentangling (e.g. TW~Dra, \cite{leh09}), allowing to recover the component spectra from the composite ones and to study the residual line profile variations{ for a spectroscopic identification of the excited modes. The results obtained from this study show} that CT~Her is a particularly interesting binary system. oEA stars indeed {provide} extremely {useful asteroseismic targets} since they offer the {opportunity} to study in detail the connection between pulsation and mass transfer in close binary systems, including the effects of tidal interaction.

\begin{acknowledgements}
Financial support from the Belgian Science Policy and the Bulgarian Academy of Sciences through the bilateral project "Astrometric, spectroscopic, 
and photometric follow-up of binary systems" (ref. BL/33/B11) is gratefully acknowledged. AS thanks Prof.~Y.~Papamastorakis, Director, and 
Dr.~I.~Papadakis for the telescope time allocated at the Observatory of Skinakas.
The data collected at BHO made use of equipment partially funded by the Belgian National Lottery (1999). This research 
made use of the SIMBAD and VIZIER databases, operated at CDS, Strasbourg, France, as well as of the ADS bibliography. 
{We furthermore thank the referee for most valuable comments.}

\end{acknowledgements}

{}


\begin{thebibliography}{}

\bibitem[Bessell 1995]{bes} Bessell, M.S., 1995, CCD Astronomy 2, No. 4, 20
\bibitem[B\'{\i}r\'o \& Nuspl 2005]{bir05} B\'{\i}r\'o, I. B., \& Nuspl, J. 2005, in: 
   'Tidal Evolution and Oscillations in Binary Stars', ASP Conf. Ser. 333, eds. A. Claret, A. Gim\'enez 
    \& J.-P. Zahn (San Francisco: ASP), 221 
\bibitem[Breger et al.\ 1993]{brg93} Breger, M., Stich, J., Garrido, R., et al.\ 1993, A\&A 271, 482....
\bibitem[Breger 2005]{brg05} Breger, M. et al.\ 2005, in: 'The Light-Time Effect in Astrophysics, Causes and Cures 
    of the O-C diagram', ASP Conf. Ser. 335, ed. C. Sterken (San Francisco: ASP), 85 
\bibitem[Castelli \& Kurucz 2003]{ca} Castelli, F. \& Kurucz, R. 2003, in IAU Symp. 20 
\bibitem[De Cat \& Aerts 2002]{dc} De Cat,P. \& Aerts,C. 2002, A\&A, 393, 965 
\bibitem[De Gr\`eve et al.\ 2009]{dgr} De Gr\`eve, J.-P., Mennekens, N., Van Rensbergen, W. \& Yungelson, L. R. 2009, in: 
    Proc. of the 8th Pacific Rim Conference on Stellar Astrophysics Conference, May 5-9, 2008, Phuket, Thailand,  
    ASP Conf. Ser. 404, eds. B. Soonthornthum, S. Komonjinda, K. S. Cheng \& K. C. Leung, 204.
\bibitem[2008a]{dd08a} Dimitrov, D., Kraicheva, Z., \& Popov, V., 2008a, IBVS 5842, 1
\bibitem[2008b]{dd08b} Dimitrov, D., Kraicheva, Z., \& Popov, V., 2008b, IBVS 5856, 1
\bibitem[2009a]{dd09a} Dimitrov, D., Kraicheva, Z., \& Popov, V., 2009a, IBVS 5883, 1
\bibitem[2009b]{dd09b} Dimitrov, D., Kraicheva, Z., \& Popov, V., 2009b, IBVS 5892, 1
\bibitem[Fitch 1976]{fit76} Fitch, W. S. 1976, in: 
    'Multiple Periodic Variable Stars', IAU Coll. 29, Sept. 1975, Budapest, Hungary, ed. W. S. Fitch (D. Reidel Publishers), 167
\bibitem[2003]{gam03} Gamarova, A. Yu., Mkrtichian, D. E., Rodr\'{\i}guez, E.,
    et al.\ 2003, in: 'Interplay of Periodic, Cyclic and Stochastic Variability in Selected Areas 
    of the H-R Diagram', ASP Conf. Ser. 292, ed. C. Sterken (San Francisco: ASP), 369
\bibitem[Gray \& Corbally 2004]{gr} Gray, R. \& Corbally, C. 2004, AJ 207, 742 \\{\it (http://www.phys.appstate.edu/spectrum/spectrum.html)} 
\bibitem[Handler et al.\ 2002]{ha} Handler, G., Balona, L. A., Shobbrook, R. R., et al. 2002, MNRAS 333, 262
\bibitem[Hoffman \& Harrison (2009)]{hoff} Hoffman, D. I. \& Harrison, T. E. 2009, in: 'Stellar Pulsation: Challenges for Theory and Observation',
        AIP Confer. Proc. 1170, 31 May -5 June 2009, Santa Fe, New Mexico, eds. J. A. Guzik \& P. A. Bradley, 429
\bibitem[2004a]{kim} Kim, S.-L., Koo, J.-R, Lee, J.A. et al.\ 2004a, IBVS 5537, 1
\bibitem[2004b]{kim1} Kim, S.-L., Lee, J. W., Kwon, S.-G. \& et al.\ 2004b, A\&A 405, 231
\bibitem[2005]{kim3} Kim, S.-L., Kim, S. H., Lee, D.-J. \& et al.\ 2005, in: 
        'Tidal Evolution and Oscillations in Binary Stars', ASP Conf. Ser. 333, eds. A. Claret, A. Gim\'enez 
        \& J.-P. Zahn (San Francisco: ASP), 217
\bibitem[2006]{kim2} Kim, S.-L., Lee, C.-U., Lee, J.W. 2006, Mem. Soc. Astron. It. 77, 184 
\bibitem[Kreiner et al.\ 2001]{kre} Kreiner, J.M. et al.\ 2001, 'An Atlas of O-C diagrams 
        of Eclipsing Binary Stars', Cracow, Poland 
\bibitem[Kurtz 2000]{kur} Kurtz D. W., 2000, in: 'Delta Scuti and Related Stars', ASP Conf. Ser. 210, Breger M. 
        \& Montgomery M. (eds.), 287
\bibitem[Kuznetsova \& Svechnikov 1990]{sk} Kuznetsova, Eh.F. \& Svechnikov, M.A. 1990, TarOT 107, 76      
\bibitem[Lampens 2006]{lam06} Lampens,P. 2006, in: 'Astrophysics of Variable Stars', 
        ASP Conf. Ser. 349, eds. Sterken, C. \& Aerts, C. (San Francisco: ASP), 153.
\bibitem[2008a]{lam08a} Lampens, P., Strigachev, A., Kim, S.-L. et al.\ 2008a, Commun. in Asteroseism. 153, 54
\bibitem[2008b]{lam08b} Lampens, P., Strigachev, A., Kim, S.-L. et al.\ 2008b, Commun. in Asteroseism. 157, 328 
\bibitem[Landolt 1992]{lan} Landolt, A.U. 1992, AJ 104, 340
\bibitem[Lehmann \& Mkrtichian 2004]{leh04} Lehmann, H.\& Mkrtichian, D. 2004, A\&A 413, 293 
\bibitem[Lehmann \& Mkrtichian 2008]{leh08} Lehmann, H.\& Mkrtichian, D. 2008, A\&A 480, 247
\bibitem[Lehmann et al.\ 2009]{leh09} Lehmann, H., Tkachenko, A., \& Mkrtichian, D. E. 2009, JENAM 2008 Symp. 4: 
        'Asteroseismology and Stellar Evolution', Vienna, Austria, Sept. 8-12, 2008, Commun. in Asteroseism. 159, eds. S. Schuh \& G. Handler, 45.
\bibitem[Lenz \& Breger 2005]{lenz} Lenz, P. \& Breger, M. 2005, CoAst 146,53
\bibitem[Maceroni 2006]{mac} Maceroni, C. 2006, in: 'Astrophysics of Variable Stars', 
        ASP Conf. Ser. 349, eds. Sterken, C. \& Aerts, C. (San Francisco: ASP), 41  
\bibitem[Mason et al.\ 2001]{mas} Mason, B.D., Gies, D.R., Hartkopf, W.I. 2001, ASSL 264, ed. D. Vanbeveren, 45
\bibitem[Mayor et al.\ 2001]{may} Mayor, M., Udry, S., Halbwachs, J.-L., et al.\ 2001, in: 'The Formation of Binary Stars', 
        IAU Symp. 200, eds. H.~Zinnecker \& R.D.~Mathieu, 47
\bibitem[Mazeh \& Zucker 1994]{maz} Mazeh, T., \& Zucker, S. 1994, Ap\&SS 212, 349
\bibitem[Michalska \& Pigulski (2008)]{mi08} Michalska, G. \& Pigulski, A. 2008, Journal of Physics: Conf. Ser. 118, 2064
\bibitem[2002]{mkr1} Mkrtichian, D.E., Kusakin, A.V., Gamarova, A.Yu. et al.\ 2002, in: 'Observational Aspects 
         of Pulsating B- and A Stars', ASP Conf. Ser. 256, C. Sterken \& D. W. Kurtz (eds.), 259 
\bibitem[2004]{mkr2} Mkrtichian, D.E., Kusakin, A.V., Rodr\'{\i}guez, E. et al.\ 2004, A\&A, 419, 1015
\bibitem[2005]{mkr3} Mkrtichian, D.E., Rodr\'{\i}guez, E., Olson, E.C. et al.\ 2005, in: 
         'Tidal Evolution and Oscillations in Binary Stars', ASP Conf. Ser. 333, eds. Claret, A., Gim\'enez, A. 
         \& J.-P. Zahn (San Francisco: ASP), 197
\bibitem[2006]{mkr4} Mkrtichian, D., Kim, S.-L., Kusakin, A. V. et al.\ 2006, Ap\&SS 304, 169
\bibitem[Mkrtichian et al.\ 2007]{mkr5} Mkrtichian,D.E., Kim,S.-L., Rodr\'{\i}guez,E. et al.\ 2007, in: 'Solar and Stellar Physics
         Through Eclipses', ASP Conf. Ser. 370, O. Demircan, S. O. Selam, \& B. Albayrak (eds.) (San Francisco: ASP), 194
\bibitem[2009]{oc} 'O-C Gateway' {\it (http://var.astro.cz/ocgate/ocgate.php)}
\bibitem[Papadakis et al.\ 2003]{pa} Papadakis, I., Boumis, P., Samaritakis, V. et al.\ 2003, A\&A 397, 565
\bibitem[Pezzullo 2011]{pez11} Pezzullo, J.~C. 2011, 'Interactive Statistics Pages'\\ {\it (http://statpages.org/pdfs.html)}
\bibitem[Pigulski \& Michalska (2007)]{pi07} Pigulski, A. \& Michalska, G. 2007, Acta Astron. 57, 61
\bibitem[Pr\v{s}a \& Zwitter 2005a]{prsa1} Pr\v{s}a,A., \& Zwitter,T. 2005a, Ap\&SS 296, 315
\bibitem[Pr\v{s}a \& Zwitter 2005b]{prsa2} Pr\v{s}a,A., \& Zwitter,T. 2005b, ApJ 628, 426 \\{\it (http://www.fiz.uni-lj.si/phoebe/)}
\bibitem[Raskin et al.\ 2011]{raskin} Raskin, G., Van Winckel, H., Hensberge, H., et al.\ 
        2011, AA 526, A69
\bibitem[Reed \& Brondel (2005)]{reed05} Reed, M. D., \& Brondel, B. J. 2005, in: 
        'Tidal Evolution and Oscillations in Binary Stars', ASP Conf. Ser. 333, eds. A. Claret, A. Gim\'enez 
        \& J.-P. Zahn (San Francisco: ASP), 228  
\bibitem[2004]{rod04} Rodr\'{\i}guez, E., Garc\'{\i}a, J.M., Mkrtichian, D.E., et al.\ 2004, MNRAS 347, 1317 
\bibitem[Rodr\'{\i}guez et al.\ 2010]{rod10} Rodr\'{\i}guez, E., Garc\'{\i}a, J.M., Costa, V., et al.\ 2010, MNRAS 408, 2149 
\bibitem[Ruc\'{\i}nski 1969]{ru} Ruc\'{\i}nski, S.M. 1969, Acta Astron. 19, 245
\bibitem[Samus et al.\ 2009]{sam} Samus, N.N., Durlevich, O.V., et al.\ 2009, Combined General Catalogue of Variable Stars,
                                 VizieR Online Data Catalog (GCVS4.2, version 2009Mar) 
\bibitem[Soydugan et al.\ 2006]{soy06} Soydugan, F., Soydugan, E., Ibanoglu, C., et al.\ 2006, AN 327, 905
\bibitem[Soydugan et al.\ (2008)]{soy08} Soydugan, F., Kacar, Y., Soydugan, E., et al.\ 2008, Commun. in Asteroseism. 157, 321 
\bibitem[Stetson 1987]{st} Stetson, P. 1987, PASP 99, 191
\bibitem[Strigachev 2009]{str} Strigachev, A. 2009, Bulg. Astron.J. 11, 87
\bibitem[Svechnikov \& Kuznetsova 1990]{sv} Svechnikov, M.A. \& Kuznetsova, Eh.F. 1990, Catalogue of approximate photometric 
                                 and absolute elements of eclipsing variable stars, VizieR Online Data Catalog (V/124)
\bibitem[Tkachenko et al.\ 2010]{tka10} Tkachenko, A., Lehmann, H. \& Mkrtichian, D. 2010, AJ 139, 1327
\bibitem[Van Hamme (1993)]{vh} Van Hamme, W. 1993, AJ 106, 2096
\bibitem[Wilson \& Devinney 1971]{wd1} Wilson, R.E. \& Devinney,E.J. 1971, ApJ 166, 605
\bibitem[1979]{wd2} Wilson, R.E. 1979, ApJ 234, 1054
\bibitem[1990]{wd3} Wilson, R.E. 1990, ApJ 356, 613
\bibitem[Witte \& Savonije 1999]{wit99} Witte, M. G. \& Savonije, G. J. 1999, A\&A 350, 129

\end{thebibliography}
\end{document}